\documentclass[12pt]{article}
\usepackage{amsmath}
\usepackage{graphicx,psfrag,epsf}
\usepackage{enumerate}
 
\usepackage{url} 

\usepackage{amsmath, amsfonts, amssymb, amstext, graphicx, epsfig, mathtools, enumerate,rotating,verbatim,subfigure}
\usepackage{bm, bbm}

\RequirePackage[authoryear]{natbib}
\RequirePackage[colorlinks,citecolor=blue,urlcolor=blue]{hyperref}

\usepackage[ruled,plain,linesnumbered]{algorithm2e} 
\makeatletter
\renewcommand{\algocf@captiontext}[2]{#1\algocf@typo. \AlCapFnt{}#2} 
\def\@algocf@capt@plain{top}
\renewcommand{\algocf@makecaption}[2]{%
  \addtolength{\hsize}{\algomargin}%
  \sbox\@tempboxa{\algocf@captiontext{#1}{#2}}%
  \ifdim\wd\@tempboxa >\hsize
    \hskip .5\algomargin%
    \parbox[t]{\hsize}{\algocf@captiontext{#1}{#2}}
  \else%
    \global\@minipagefalse%
    \hbox to\hsize{\box\@tempboxa}
  \fi%
  \addtolength{\hsize}{-\algomargin}%
}
\makeatother

\usepackage{float}
\usepackage{prettyref,soul}

\usepackage{tabulary}
\usepackage[usenames]{color}
\usepackage{multirow}
\usepackage{tabularx}
\usepackage{booktabs}
 \usepackage{url}
\usepackage{bbding}
\usepackage{mathrsfs}

\usepackage[normalem]{ulem}  
\usepackage[table]{xcolor}  
\usepackage{xcolor}  

\newrefformat{eq}{(\ref{#1})}
\newrefformat{chap}{Chapter~\ref{#1}}
\newrefformat{sec}{Section~\ref{#1}}
\newrefformat{algo}{Algorithm~\ref{#1}}
\newrefformat{fig}{Fig.~\ref{#1}}
\newrefformat{tab}{Table~\ref{#1}}
\newrefformat{rmk}{Remark~\ref{#1}}
\newrefformat{clm}{Claim~\ref{#1}}
\newrefformat{def}{Definition~\ref{#1}}
\newrefformat{cor}{Corollary~\ref{#1}}
\newrefformat{lmm}{Lemma~\ref{#1}}
\newrefformat{lemma}{Lemma~\ref{#1}}
\newrefformat{prop}{Proposition~\ref{#1}}
\newrefformat{app}{Appendix~\ref{#1}}
\newrefformat{ex}{Example~\ref{#1}}
\newrefformat{exer}{Exercise~\ref{#1}}
\newrefformat{soln}{Solution~\ref{#1}}
\newrefformat{cond}{Condition~\ref{#1}}

\newtheorem{theorem}{Theorem}[section]
\newtheorem{lemma}{Lemma}[section]
\newtheorem{proposition}[theorem]{Proposition} 
\newtheorem{remark}{Remark}[section]

\newtheorem{assumption}{Assumption}[section]

\def\text#1{\mbox{\rm #1}}
\def\overset#1#2{\stackrel{#1}{#2} }

\def\dfrac{\displaystyle\frac}

\newcommand{\ie}{\mbox{\sl i.e.\;}}
\newcommand{\eg}{\mbox{\sl e.g.\;}}

\newcommand{\wh}{\hat}
\renewcommand{\hat}{\widehat}

\def\be{\begin{align}}
\def\ee{\end{align}}
\def\T{{ \mathrm{\scriptscriptstyle T} }}

\newcommand{\Dtest}{{\cal D}_{\rm test}}

\newcommand{\Itesti}{{\cal I}_{\rm test}^{i_0}}
\newcommand{\Itraini}{{\cal I}_{\rm train}^{i_0}}
\newcommand{\Icalibi}{{\cal I}_{\rm calib}^{i_0}}

\newcommand{\cR}{\mathcal{R}}

\newcommand{\alphaBH}{\alpha_{\mathrm{BH}}}
\newcommand{\alphaeBH}{\alpha_{\mathrm{eBH}}}
\newcommand{\td}{{\mathrm{d}}}

\newrefformat{eq}{(\ref{#1})}
\newrefformat{chap}{Chapter~\ref{#1}}
\newrefformat{sec}{Section~\ref{#1}}
\newrefformat{algo}{Algorithm~\ref{#1}}
\newrefformat{fig}{Fig.~\ref{#1}}
\newrefformat{tab}{Table~\ref{#1}}
\newrefformat{rmk}{Remark~\ref{#1}}
\newrefformat{clm}{Claim~\ref{#1}}
\newrefformat{def}{Definition~\ref{#1}}
\newrefformat{cor}{Corollary~\ref{#1}}
\newrefformat{lmm}{Lemma~\ref{#1}}
\newrefformat{lemma}{Lemma~\ref{#1}}
\newrefformat{prop}{Proposition~\ref{#1}}
\newrefformat{app}{Appendix~\ref{#1}}
\newrefformat{ex}{Example~\ref{#1}}
\newrefformat{exer}{Exercise~\ref{#1}}
\newrefformat{soln}{Solution~\ref{#1}}
\newrefformat{cond}{Condition~\ref{#1}}

\def\text#1{\mbox{\rm #1}}
\def\overset#1#2{\stackrel{#1}{#2} }

\def\dfrac{\displaystyle\frac}
 
\def\be{\begin{equation}}
\def\ee{\end{equation}}
\def\T{{ \mathrm{\scriptscriptstyle T} }}

\newcommand{\bA}{{\mathbf A}}

\newcommand{\bM}{{\mathbf M}}

\usepackage{float}
\usepackage{setspace}

\newcommand{\newtext}[1]{\textcolor{black}{#1}}

\graphicspath{{./figs/}}
\usepackage[margin=1in]{geometry}

\begin{document}

\def\spacingset#1{\renewcommand{\baselinestretch}%
{#1}\small\normalsize} \spacingset{1}


\title{\bf Conformal network link prediction with false discovery rate control under unstructured missingness}
\author{
Wenqin Du\textsuperscript{1}\thanks{Equal contribution.},
Wanteng Ma\textsuperscript{2}\footnotemark[1],
Dong Xia\textsuperscript{3},
Yuan Zhang\textsuperscript{4}\thanks{Corresponding author: \url{mailto:yzhanghf@stat.osu.edu}},
and Wen Zhou\textsuperscript{5}
}
\date{
University of Southern California\textsuperscript{1},
University of Pennsylvania\textsuperscript{2},\\
Hong Kong University of Science and Technology\textsuperscript{3},\\
Ohio State University\textsuperscript{4},
and New York University\textsuperscript{5}
}
\maketitle

\begin{abstract}
We propose a new method for predicting multiple missing links in partially observed networks while controlling the false discovery rate (FDR), a largely unresolved challenge in network analysis. 
The main difficulty lies in handling complex dependencies and unknown missing patterns. 
We introduce conformal link prediction, a distribution-free procedure grounded in the exchangeability structure of weighted graphon models. 
Our approach constructs conformal p-values via a novel multi-splitting strategy that restores exchangeability within local test sets, thereby ensuring valid row-wise FDR control, even under unknown missing mechanisms. 
To achieve FDR control across all missing links, we further develop a new aggregation scheme based on e-values, which accommodates arbitrary dependence across network predictions. 
Our method requires no assumptions on the missing rates, applies to weighted, unweighted, undirected, and bipartite networks, and enjoys finite-sample theoretical guarantees. 
Extensive simulations and real-world data study confirm the effectiveness and robustness of the proposed approach.
\end{abstract}


\spacingset{1.45} 

\section{Introduction}
\label{section::intro}


Link prediction \citep{janicik2005social,clauset2008hierarchical,kim2011network,mohan2021graphical} is a fundamental problem in network data analysis, with applications in sociology, econometrics, and biology, where real-world networks are often only partially observed.
While early work focused on accurately predicting individual links, the growing demand for simultaneous prediction of multiple links has brought concerns about reproducibility and statistical validity to the forefront. Framing this task as a multiple hypothesis testing problem has led to the adoption of false discovery rate (FDR) control as a key tool. However, FDR control for link prediction in networks presents unique challenges owing to complex edge dependencies and unknown, heterogeneous patterns of missingness. To address these challenges, which remain inadequately resolved in the literature, we introduce conformal link prediction ({\tt clp}), a new method for simultaneous link prediction with provably FDR control.

Let $\bA \in \mathbb{R}^{n\times n}$ denote the adjacency matrix of a directed network with $n$ nodes and no self-loops, where entries $A_{i,j}$ and $A_{j,i}$ are not necessarily equal. Missingness in $\bA$ is described by a binary matrix $\bM\in \{0,1\}^{n\times n}$, where its entry $M_{i,j}=1$ if $A_{i,j}$ is missing, {and $\bM$ can be any deterministic or random binary matrix.} We define $\bM^c=\boldsymbol{1}_{n\times 1}\boldsymbol{1}_{n\times 1}^{\T}-\bM$. Throughout, we only adopt the following standard assumption.
\begin{assumption}[\citet{shao2023distribution}]
    \label{assumption::rm}
    The  missing pattern $\bM$  is independent of the network, specified by the adjacency matrix $\bA$.
\end{assumption} 

\newtext{
Assumption \ref{assumption::rm} is closely related to the classical missing-completely-at-random (MCAR) condition in the missing data literature \citep{rubin1976inference,molenberghs2014handbook,little2019statistical}. In our graphon setting, it requires the missingness pattern $\bM$ to be independent of the realized network $\bA$, while still allowing heterogeneous or structured missing patterns. This differs from missing-at-random (MAR) mechanisms, where missingness may depend on the latent node variables $\{\xi_i\}_{i=1}^n$ but not directly on the realized link values conditional on these latent variables. It also excludes missing-not-at-random (MNAR) mechanisms, where missingness may depend directly on the unobserved value $A_{i,j}$. 
}

We define the test set as the collection of unobserved links:
$\Dtest:=\{(i,j): i,j \in [n]=\{1,\ldots, n\}, M_{i,j}=1\}$, 
where $\Dtest$ is random if $\bM$ is random. For simplicity of presentation, we focus on predicting all links in $\Dtest$, while our method can be applied to any subset. 
Following \citet{jin2022selection}, we formulate link prediction as a multiple testing problem with random hypotheses:
\begin{equation} 
    H_{i,j}: A_{i,j} \leq c_{i,j}~ 
    \text{for } (i,j) \in \Dtest,
    \label{def::hypothesis-testing}
\end{equation} where $c_{i,j}$'s are user-specified. For example, in the international trade flow network \citep{helpman2008estimating}, if the goal is to determine whether any trading activity exists between two countries, $c_{i,j}$ can be set to zero. Alternatively, if the focus is on detecting higher volumes of bilateral trade, $c_{i,j}$ can be set to an appropriate positive value. 

For the rejection set $\cR:=\{{(i,j) \in \Dtest}: \textrm{Reject }H_{i,j}\}$, the false discovery proportion (FDP) is defined as $\textrm{FDP}=
(|\cR| \vee 1)^{-1} \sum_{(i,j)\in \cR} \mathbbm{1}_{[A_{i,j}\leq c_{i,j}]}$, based on given $\{c_{i,j}\}_{(i,j) \in \Dtest}$ and observed data $\big\{\bA\circ \bM^c, \bM\big\}$, where ``$\circ$'' denotes the element-wise product. {Our objective is to control the conditional FDR given $\bM$ at a pre-specified significance level, and accordingly treating the test set as fixed. Clearly, the marginal FDR control is thus also valid since one can marginalize the conditional FDR control guarantee over $\bM$. Specifically, we define $\textrm{FDR}=\mathbb{E}(\textrm{FDP}|\bM)$.} A major challenge for FDR control in classical link prediction methods is that the unknown and potentially non-uniform missing patters lead to non-exchangeable observations $\bA\circ \bM^c$, which hinders uncertainty quantification. Moreover, complex dependencies among predicted links further render most traditional FDR control procedures inapplicable.  

\subsection{Related literature}

Link prediction has been extensively studied over recent decades \citep{liben2003link,hasan2011survey, zhang2017estimating}. 
An alternative perspective treats the network as a random matrix or a realization of an unobservable graphon, leading to approaches based on matrix estimation or completion \citep{candes2011robust,koltchinskii2011nuclear,davenport20141} and graphon estimation \citep{gao2015rate,xu2018rates}. 
However, most existing studies either lack uncertainty quantification or rely on strong modeling assumptions. 
To achieve our goal of distribution-free uncertainty quantification, we bridge two key ingredients.
The first is conformal prediction \citep{vovk2005algorithmic},
a powerful model-agnostic framework that enables flexible applications,  including recent adaptations to multiple testing via the construction of conformal $p$-values \citep{lei2021conformal, jin2022selection, bates2023testing}.
The second ingredient is the e-BH procedure \citep{wang2022false}, which ensures valid FDR control under complex dependence among conformal $p$-values. Based on e-values \citep{vovk2021values}, this procedure has proven effective in handling complex dependence among statistics, with recent successful applications in derandomized knockoffs \citep{ren2024derandomised} and e-value-based conformal prediction \citep{bashari2024derandomized,ignatiadis2024values}.

Conformal prediction for matrix data was pioneered by \cite{gui2023conformalized} and \citet{shao2023distribution}. 
\cite{gui2023conformalized} considered random missingness and is therefore more closely related to our setting, whereas \citet{shao2023distribution} examined arbitrary and potentially adversarial missing patterns.
However, neither study addresses the task of predicting multiple links. Recently, \citet{marandon2024conformal} introduced conformal link prediction with FDR control, but the method is limited to unweighted networks and assumes a ``double-standard" sampling scheme, requiring uniform missing rates across all links. Under this assumption, subsequent work by \cite{blanchard2024fdr} sought to provide theoretical justification for \citet{marandon2024conformal}.
This uniformity condition effectively imposes the exchangeability on all links with values exceeding $c_{i,j}$, simplifying the analysis but limiting practical applicability.
\citet{liang2024structured} proposed a structured splitting scheme for simultaneous prediction of a small number of links, without considering the multiple testing.
{\cite{zhao2024conformalized} construct prediction intervals for individual links with marginal coverage guarantees, rather than controlling the FDR for multiple dependent links simultaneously.}
To date, valid FDR control for multiple link prediction under general network settings remains an open challenge. 

\subsection{Our contributions}

We develop a new conformal inference framework for multiple network link prediction with provable FDR control. Unlike prior work, our method is tailored to the complex structure of partially observed networks, characterized by heterogeneous missingness and intricate dependencies among predictions. The approach combines novel conformal calibration schemes with a principled global aggregation strategy. 
Our main contributions are threefold.

First, we introduce a conformal inference method tailored to row-wise link prediction.
A key innovation is a carefully constructed multi-splitting scheme that partitions each row's test set and adaptively allocates calibration sets. 
This enables the construction of exchangeable nonconformity scores and valid conformal $p$-values, even under unknown and nonuniform missingness. 
The method achieves FDR control for any node sizes $n$ through a nontrivial adaptation of the Benjamini–Hochberg (BH) procedure \citep{benjamini1995controlling}, applied in a row-wise fashion.

Second, we propose a novel aggregation technique that transforms the row-wise test decisions, obtained via the BH procedure, into valid e-values, and applies the e-BH procedure for FDR control. 
This step requires careful design to accommodate unknown yet arbitrary dependence across predicted network links, a challenge not addressed by standard conformal or e-value methods. We further introduce a derandomization step to enhance stability, along with an e-value inflation strategy to improve power, both tailored to the unique demands for predicting multiple network links.

Third, our method is broadly applicable: it extends naturally to unweighted, undirected, and bipartite networks, requires no knowledge of the missingness mechanism, and provides finite-sample guarantees. 
Our framework advances the scope of conformal inference in network analysis, extending beyond existing approaches such as \citet{tibshirani2019conformal} and \citet{gui2023conformalized}. Numerical studies confirm the practical effectiveness of the proposed method.

\section{Conformal link prediction with FDR control}
\label{section::cap-with-fdr-control}

\subsection{Link prediction under unknown missing probabilities}
\label{section::2-1}

We adopt the general weighted graphon model \citep{gao2016optimal} as the base for our method for its flexibility and generality. 
We assume entry $A_{i,j}$ of the adjacency matrix $\bA$ is drawn from a distribution parameterized by $f(\xi_i,\xi_j)$, such as Bernoulli and Gaussian. That is, \begin{align}
    A_{i,j} 
    \sim
    f(\xi_i,\xi_j), \text{ ~for~ } i\neq j \in [n],
    ~~
    \text{where }\{\xi_i\}_{1\leq i\leq n} \stackrel{\rm i.i.d.} \sim \mathrm{Uniform}[0,1].
    \label{def::exch-model}
\end{align} Intuitively, model \eqref{def::exch-model} first samples nodes independently from a hyper-population of ``all kinds of nodes", then generates links according to the function $f$. This framework encompasses a broad class of network models, such as in the social relations model \citep{kenny1984srm}, the conditionally independent dyad model \citep{graham2020cid}, and the random-effects model \citep{gelman2006ref}. 

Under model \eqref{def::exch-model}, for any permutation $\pi(\cdot)$ of node labels, we have 
\begin{equation}
    \{A_{i,j}\}_{i\neq j \in [n]}
    \overset{d}{=}
    \{A_{\pi(i),\pi(j)}\}_{i\neq j \in [n]},
    \label{def::exch-property}
\end{equation}
which is known as the exchangeability property of $\bA$. Importantly, the exchangeability in \eqref{def::exch-property} pertains to the not-fully-observable complete $\bA$, not the observed $\bA\circ \bM^c$.
Note that our notion of exchangeability in \eqref{def::exch-property} differs from that in prior work. 
Previous studies \citep{tibshirani2019conformal,gui2023conformalized} rely on  exchangeability induced by the random $\bM$ where they assume $M_{i,j}\sim\operatorname{Ber}(q_{i,j})$, and therefore require either uniform missing rates ${q_{i,j}}$ or their accurate estimations under the ``weighted exchangeability" framework.
In contrast, we model $\bA$ as a random graph with inherent exchangeability as defined in \eqref{def::exch-property}. 
With the mild additional assumption of exchangeability of the complete $\bA$, we waive the need for prior knowledge of the potentially non-uniform missing rates ${q_{i,j}}$'s, a key restrictive assumption in earlier works.

To address the challenge of unknown missing patterns, we propose a divide-and-conquer approach to break the problem into sub-tasks.
First, for each row $i_0$, we construct ``local tests'' where exchangeability holds within a row-wise subset of observations, which leads to valid ``local'' FDR controls. 
Then, the local test decisions are aggregated into e-values to construct an aggregated test of the hypotheses in \eqref{def::hypothesis-testing} for all unobserved links, with valid FDR control.

\subsection{Step 1: breaking down original test into smaller local tests}
\label{section::2-2}

\begin{figure}[h]
    \centering
    \includegraphics[width=0.9\textwidth]{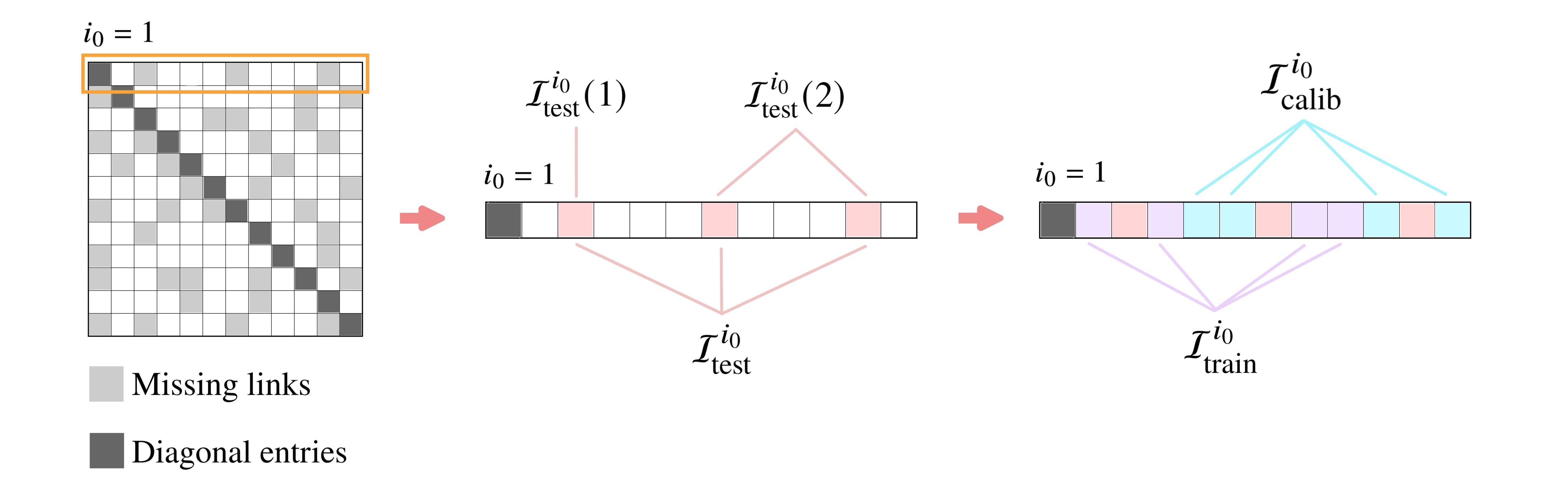}
    \caption{
    Illustration of $\Itesti$, $\Itraini$, and $\Icalibi$ within row $i_0$. {Here, the local test set $\Itesti$ is split into two subsets for the divide-and-conquer approach, namely $\Itesti(1)$ and $\Itesti(2)$.}
    }
    \label{fig::row-illus-ex}
\end{figure}

Our first step is to design a {\it local test}, which focuses only on row $i_0$, by constructing exchangeable conformal $p$-values \citep{jin2022selection}. 
Let $\Itesti=\{(i_0,j): j\in [n], M_{i_0,j}=1\}$ denote the column indices of missing entries in row $i_0$, corresponding to missing links for node $i_0$.
Then, the corresponding set of local hypotheses is
\begin{equation}
    H_{i_0,j}: A_{i_0,j} \leq c_{i_0,j},
    ~~
    \text{for }j \in \Itesti.
    \label{def::hypothesis-testing-local}
\end{equation}
We randomly split the observed entries in row $i_0$, that is, 
$\{A_{i_0,j} : j \notin \Itesti \}$, into a training set $\{A_{i_0,j} : j \in \Itraini \}$ and a calibration set $\{A_{i_0,j} : j \in \Icalibi \}$, as illustrated in Figure \ref{fig::row-illus-ex}.

Under the random missing mechanism in Assumption \ref{assumption::rm}, the observed entries in the same row can be viewed as a randomly selected subset of exchangeable random variables.
It turns out that, as a consequence of
the exchangeability of latent factors $\{\xi_i\}_{1\leq i\leq n}$, the entries in $\{A_{i_0,j}\}_{j \in \Icalibi \cup \Itesti}$ are exchangeable, as established in the following proposition.
\begin{proposition}
\label{prop::withinrow-exch}
Under Assumption \ref{assumption::rm} and model  \eqref{def::exch-model}, given $\bM$, for any $i_0 \in [n]$ and any permutation $\widetilde\pi(\cdot)$ of index set $\Icalibi \cup \Itesti$, we have 
$\{A_{i_0,j}: j \in \Icalibi \cup \Itesti\}\overset{d}{=}\{A_{i_0,\widetilde\pi(j)}: j \in \Icalibi \cup \Itesti\}$.
\end{proposition}

For each $j \in \Icalibi \cup \Itesti $, we select data from observed links using a carefully designed strategy, detailed later in Section \ref{sec23}, to construct a link prediction $\wh A_{i_0,j}$ such that 
\begin{align}
    \{\wh A_{i_0,j}\} \text{ are exchangeable for } j\in \Icalibi \cup\Itesti.
  \label{eq:est-cond}
\end{align}
Based on \eqref{eq:est-cond}, we construct the 
nonconformity scores \citep{jin2022selection} as follows $S(\wh A_{i_0,j}; z) := z - \wh A_{i_0,j}$ for $j \in \Icalibi \cup \Itesti$, where
where $S(\wh A_{i_0,j}; z)$ is monotone with respect to the second argument by definition. Intuitively, $S(\wh A_{i_0,j}; z)$ quantifies how well the value $z$ conforms to the predictions. Using the nonconformity scores, for each $j_1 \in \Itesti$, we define $p$-values for entries $\{(i_0,j_1):j_1 \in \Itesti\}$ as
$\wh{\mathfrak p}_{i_0,j_1}=
(1 + |\Icalibi|)^{-1}[1 +\sum_{j' \in \Icalibi}\mathbbm{1}\{ S(\wh A_{i_0,j'}; A_{i_0,j'}) < S(\wh A_{i_0,j_1}; c_{i_0,j_1}) \}]$.
We then apply the BH procedure to  
$\big\{\wh{\mathfrak p}_{i_0,j}: j \in \Itesti \big\}$ as follows. Let $\wh{\mathfrak p}_{(\ell)}^{i_0}$ denote the $\ell$th smallest element of the sorted values of 
$\big\{\wh{\mathfrak p}_{i_0,j}: j \in \Itesti
\big\}$. Given any $\alphaBH>0$, define
\begin{align}
    \wh\ell
    :=
    \max\big\{
        \ell:
        \wh{\mathfrak p}_{(\ell)}^{i_0}
        \leq
        \alphaBH \cdot \ell / |\Itesti|
    \big\}.
    \label{equation::BH-threshold-l}
\end{align}
We reject $H_{i_0,j}$ for all $j$ such that $\wh{\mathfrak p}_{i_0,j} < \wh{\mathfrak p}_{(\wh\ell)}^{i_0}$, and denote the rejection set by $\cR_{\rm BH}^{i_0}(\alphaBH)$. The following result establishes FDR control for this local testing procedure.  
\begin{theorem}[Oracle FDR control for the local test with conformal $p$-values] 
\label{thm::FDR-rowwise}
    For network $\bA$ generated from \eqref{def::exch-model}, suppose the prediction $\widehat \bA$ satisfies \eqref{eq:est-cond} and $\{A_{i_0,j} - \wh A_{i_0,j}\}_{j \in \Icalibi} \cup \{c_{i_0,j} - \wh A_{i_0,j}\}_{j \in \Itesti}$ have no ties almost surely. 
    Then, for any $\alphaBH \in (0,1)$, {conditional on $\bM$}, the rejection set $\cR_{\rm BH}^{i_0}(\alphaBH)$ 
    for testing local hypotheses in \eqref{def::hypothesis-testing-local} satisfies FDR $\leq \alphaBH$.
\end{theorem}  
\begin{remark}
In presence of ties, one can simply apply a random tie-breaking trick as that in \citet{lei2018distribution,tibshirani2019conformal}.
\end{remark}

\subsection{Construction of $\widehat \bA$ and local tests with FDR control}
\label{sec23}
{To build an operational method, we construct $\widehat \bA$ that satisfies \eqref{eq:est-cond} for implementing the local test in Section~\ref{section::2-2}.}
Due to the unknown missingness, directly applying popular existing link prediction methods, such as matrix factorization or node similarity-based tools, would fail the required
exchangeabililty condition in \eqref{eq:est-cond}.
To better illustrate our strategy, we start with an ``oracle method''.
Consider a weighed link estimation:
\begin{equation}
\label{eq:weighted-est}
    \wh  A_{i_0,j_1} =  \sum\nolimits_{j_2 \in \Itraini} w(j_1,j_2)
    A_{i_0,j_2}
\end{equation}
for $j_1\in \Icalibi \cup\Itesti$,
where the weight $w(j_1,j_2)$ may be data-dependent. 
Inspired by \cite{zhang2017estimating}, for any $j_1 \in \Itesti \cup \Icalibi$ and $j_2 \in \Itraini$, a common  choice of $w(j_1,j_2)$ is
\begin{align}
    w(j_1,j_2)
    = 
    \frac{K\big(d_{j_1,j_2}\big)}
{\sum_{j_2' \in \Itraini} K\big(d_{j_1,j_2'}\big)},
    \label{method::A-estimate-oracle}
\end{align}
where $K(\cdot)$ is a smooth and Lipschitz kernel function, and ${d}(j_1,j_2)    := \int_0^1 \Big|\int_0^1 \big\{f(u,\xi_{j_1})-f(u,\xi_{j_2})\big\} f(u,v) \td u \Big| \td v$
measures the dissimilarity of graphon slices $f(\cdot, \xi_{j_1})$ and $f(\cdot, \xi_{j_2})$. Since the weights $ w(j_1,j_2)$ are independent of $\bM$, the unknown missing pattern does not affect the exchangeability of the entries in $\{A_{i_0,j} - \widehat A_{i_0,j}: j \in \Icalibi \cup \Itesti\}$.
As a result, the validity of FDR control for the local test via the downstream BH procedure follows naturally from Theorem~\ref{thm::FDR-rowwise}.


To make \eqref{eq:weighted-est} operational in practice, we replace the oracle dissimilarity $d$ in \eqref{method::A-estimate-oracle} with a computable $\widehat d$.
Noted by \citet{shao2023distribution}, for each $j$, we can view $A_{i_0,j}$ as the response and $A_{(-i_0),j}$, the remainder of column $j$, as the predictor. The challenge, however, lies in constructing a suitable surrogate for the unobservable graphon slices $f(\cdot, \xi_j)$'s in \eqref{method::A-estimate-oracle}, such that $\widehat d$ would not break exchangeability -- for instance, simply filling the unobserved entries with 0 and plug in Equation (9) of \citet{zhang2017estimating} will not work.

Our key idea is to select a carefully chosen subset of $[\bA\circ \bM^c]_{(-i_0),j}$'s that maintains exchangeability.
This is another layer of divide-and-conquer approach.
Rather than using the entire calibration set $\Icalibi$ to test all hypotheses in $\Itesti$, we divide $\Icalibi$ into $|\Itesti|$ smaller subsets, and randomly assign one subset $\Icalibi(j_0)$ to each $j_0\in\Itesti$. 
\newtext{However, if $|\Itesti|$ is not small relative to $|\Icalibi|$, there may be insufficient calibration samples. In such nontrivial cases, we partition $\Itesti$ into disjoint subsets $\{\Itesti(1), \Itesti(2), \ldots, \Itesti(k)\}$ and reuse the same calibration set $\Icalibi$ across each subset $\Itesti(k_0)$, for $k_0 \in [k]$.}  The results are then aggregated using the e-BH procedure to form an aggregated test, as detailed in Section \ref{section::2-3}. For the remainder of this paper, we focus on the nontrivial case where $|\Itesti|$ is not small compared to $|\Icalibi|$. We therefore test hypotheses within a partition $\Itesti(k_0)$: 
\begin{align}
    H_{i_0,j}: A_{i_0,j} \leq c_{i_0,j},
    ~~
    \text{where }j \in \Itesti(k_0).
    \label{def::hypothesis-testing-local-sampleV}
\end{align}
Let $r_0$ denote the minimum size of each calibration split $\Icalibi(j_0)$. We choose the partition so that $|\Itesti(k_0)|\leq r_1:= \lfloor |\Icalibi|/r_0 \rfloor$, which requires $\lceil|\Itesti|/r_1\rceil$ partitions. 
\newtext{
Consequently, provided that $|\Icalibi|\ge r_0$, the calibration set can be partitioned within each local test so that every test point is assigned a calibration subset of size at least $r_0$.
}
\begin{figure}[ht]
    \centering
    \includegraphics[width=0.85\textwidth]{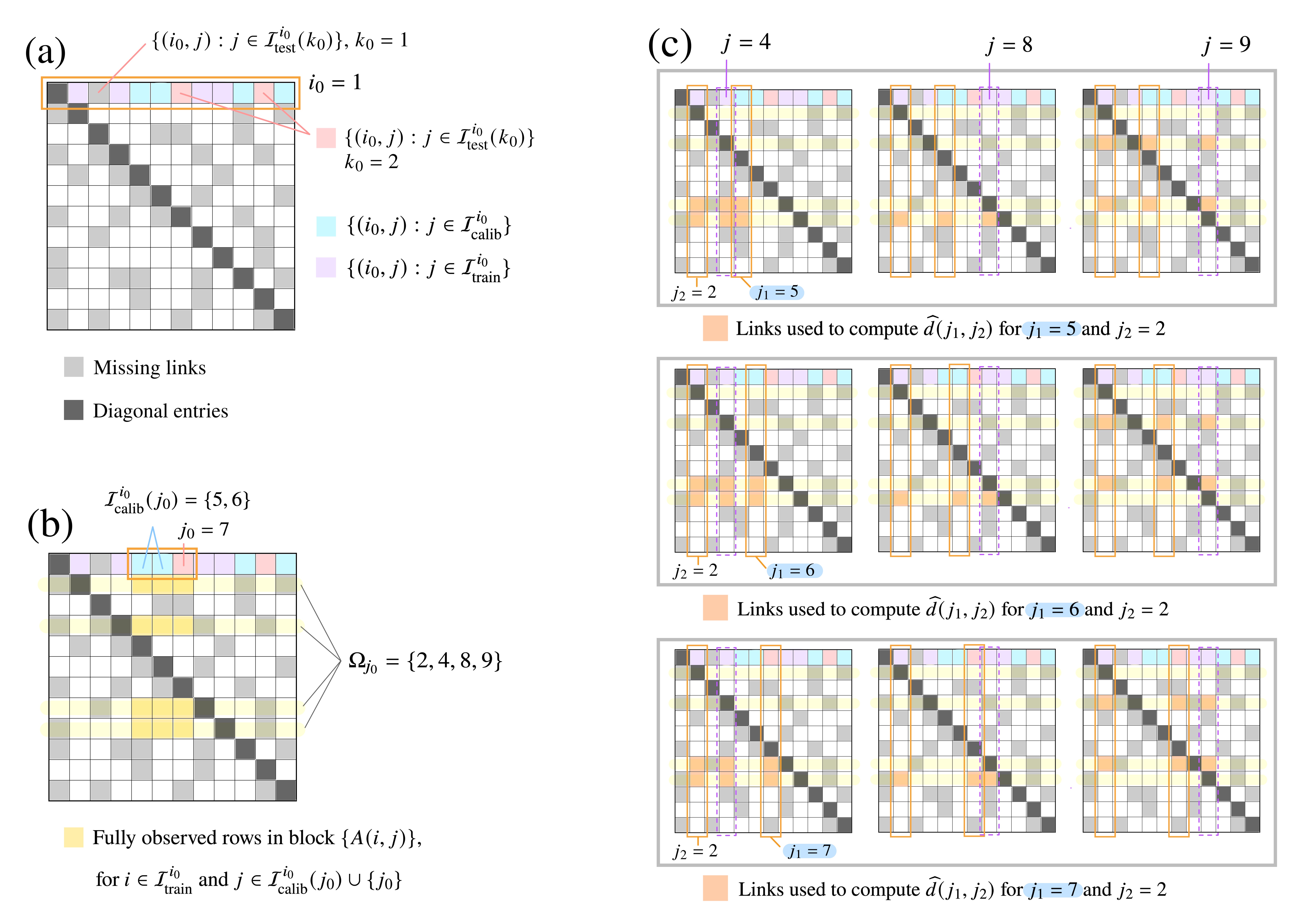}
    \caption{
    Construction of $\widehat A_{j_0,j_1}$.
    (a) Splitting row $i_0$ into training and calibration sets.
    (b) Construction of $\Omega_{j_0}$. 
    (c) Dissimilarity measures. 
    }
    \label{fig::clp-alg-illus-ex}
\end{figure}

With this setup, we illustrate the core of our method in Figure \ref{fig::clp-alg-illus-ex}.
Recall that row $i_0$ is partitioned into training, calibration, and test sets, as shown in Figure \ref{fig::clp-alg-illus-ex}-(a), {and this partition aligns with the splitting strategy illustrated in Figure~\ref{fig::row-illus-ex}}.
Figure \ref{fig::clp-alg-illus-ex}-(b) presents a simplified case where the calibration set is $\Icalibi(j_0)=\{5,6\}$, and the test set contains a single entry $\{j_0\}:=\{7\}$.
When $|\Icalibi(j_0) \cup \{j_0\}|$ is small, with a high probability, we can identify multiple rows $i\in \Itraini$, 
collectively denoted as $\Omega_{j_0}$, for which the submatrix $\bA_{\Omega_{j_0}\times (\Icalibi(j_0) \cup \{j_0\})}$ contains no missing entries, highlighted in dark yellow in Figure \ref{fig::clp-alg-illus-ex}-(b).

Note that the construction of $\Omega_{j_0}$ depends on the missing pattern. {If $\Omega_{j_0}=\emptyset$, we can simply define $\widehat{d}_{j_1,j_2,j}=1$ and thus $\widehat{d}_{j_1,j_2}=1$. It is clear that the exchangeability still holds for  $\widehat{A}_{i_0,j}$ under such uniform weight construction.}

Next, we construct $\widehat A_{i_0, j_1}$ for all $j_1\in \Icalibi(j_0) \cup \{j_0\}$.
Recall that 
we estimate $A_{i_0,j_1}$ in \eqref{eq:weighted-est} using a weighted average of $A_{i_0,j_2}$'s for $j_2\in \Itraini$, where the weights depend on a dissimilarity measure $\widehat d_{j_1,j_2}$. We now formally define $\widehat d_{j_1,j_2}$. 
For each $j\in \Itraini\backslash\{j_2\}$, let $\Omega_{j_1,j_2,j}$ be the largest subset of $[n]$ satisfying: (1) $\Omega_{j_1,j_2,j}\subseteq \Omega_{j_0}$; and (2) the submatrix $A_{\Omega_{j_1,j_2,j}\times \{j_1,j_2,j\}}$ contains no missing entries.
Let $\langle\cdot,\cdot\rangle$ denote the inner product of two vectors. We then define $\widehat d_{j_1,j_2,j}$ and use it to construct $\widehat d_{j_1,j_2}$, as follows:
\allowdisplaybreaks
\begin{align}
    \widehat d_{j_1,j_2,j}
    :=&~
   \frac{\big|
        \langle
            A_{\Omega_{j_1,j_2,j}, j_1} - A_{\Omega_{j_1,j_2,j}, j_2},
            A_{\Omega_{j_1,j_2,j}, j}
        \rangle
    \big|}{|\Omega_{j_1,j_2,j}|},
    \label{def::d_{j_1,j_2,j}}
    \\
    \text{and}\quad
    \widehat d_{j_1,j_2}
    :=&~
    \sum\nolimits_{j \in \Itraini \backslash \{j_2\}} \frac{\widehat d_{j_1,j_2,j}}{|\Itraini| - 1}.
    \label{def::d_{j_1,j_2}}
\end{align}
Then similar to \eqref{method::A-estimate-oracle}, we can construct, for $j_1\in \Icalibi(j_0)\cup\{j_0\}$,
\begin{equation}    \label{method::A-estimate} 
\wh A_{i_0,j_1}:=
\left\{\sum\nolimits_{j_2' \in \Itraini} K\big(\wh d_{j_1,j_2'}\big)\right\}^{-1}{\sum\nolimits_{j_2' \in \Itraini} K\big(\wh d_{j_1,j_2'}\big) A_{i_0,j_2'}}.
\end{equation} For ease of reference, Table 1 summarizes the index notation introduced thus far.


Let us explain why the $\widehat A_{i_0,j_1}$'s constructed in such a way are exchangeable.
By Proposition \ref{prop::withinrow-exch}, it suffices to show that the dissimilarities $\widehat d_{j_1,j_2}$'s are exchangeable for all $j_1\in \Icalibi(j_0) \cup \{j_0\}$, which 
in turn follows from the exchangeability of the underlying $\widehat d_{j_1,j_2,j}$'s for these $j_1$'s.
Now we make a crucial observation that by construction, for each fixed pair $(j_2, j)$, all indices $j_1\in \Icalibi(j_0) \cup \{j_0\}$ share a common $\Omega_{j_1,j_2,j}$.
For clarity, this observation is illustrated by comparing the three vertically stacked panels in Figure \ref{fig::clp-alg-illus-ex}-(c), where dark orange cells highlight the rows corresponding to $\Omega_{j_1,j_2,j}$. By Assumption \ref{assumption::rm} and model \eqref{def::exch-model}, conditional on $\bM$ and $\{\xi_{j'}: j'\in \Itraini\}$, $\xi_{j_1}$'s are independent and identically distributed for $j_1\in \Icalibi(j_0) \cup \{j_0\}$. This immediately implies the exchangeability of $\widehat d_{j_1,j_2,j}$'s, and hence that of $\widehat A_{i_0,j_1}$'s, for all $j_1\in \Icalibi(j_0) \cup \{j_0\}$.
Consequently, the nonconformity scores $S(\wh A_{i_0,j'}; A_{i_0,j'})$ for $j'\in \Icalibi(j_0)\cup\{j_0\}$ are also exchangeable. Therefore, for $j_0 \in \Itesti(k_0)$,
\begin{equation}\label{eqn-def::emp-p-value}
    \begin{aligned}
        \wh{\mathfrak p}_{i_0,j_0}
        :=&~
        \dfrac
            {
                1 +
                \sum_{j' \in \Icalibi(j_0)}
                \mathbbm{1}\{ S(\wh A_{i_0,j'}; A_{i_0,j'}) < S(\wh A_{i_0,j_0}; c_{i_0,j_0})\}
            }
            {
                1 + |\Icalibi(j_0)|
            }
    \end{aligned}
\end{equation}
are valid conformal $p$-values \citep{jin2022selection,bates2023testing}. Moreover, they are conditionally independent by Proposition \ref{prop:emp-cond-ind}, making them suitable for the BH procedure.

\begin{table}[]
    \label{tab:summary-index}
    \centering
    \begin{tabular}{|c|l|}
    \hline
      Index &  Meaning \\
      \hline
      $i_0,j_0$ &  position of missing entry to construct $\wh{\mathfrak p}_{i_0,j_0}$ \\
      \hline
       $j_1$ & element in calibration set: $j_1\in \Icalibi(j_0) \cup \{j_0\}$ \\
       \hline
       $j_2$ & element in training set: $j_2\in \Itraini$ \\
      \hline
        $j,j'$  & auxiliary indices for summation  \\
       \hline
        $k_0$ & subset that $j_0$ belongs to: $j_0 \in \Itesti(k_0)$\\
         \hline
    \end{tabular}
    \caption{A summary of index notation used in this section}
\end{table}

\begin{proposition}[Conditional independence of empirical $p$-values]\label{prop:emp-cond-ind}
For network $\bA$ generated from \eqref{def::exch-model}, under Assumption \ref{assumption::rm}, conditional on 
$\{\xi_j:j\in\Itraini\}$, $\{A_{i,j}:i,j\in\Itraini\}$ and the missing pattern $\bM$, the $p$-values in \eqref{eqn-def::emp-p-value} are independent within $\Itesti(k_0)$. 
\end{proposition}

Building on the above property of $\wh{\mathfrak p}_{i_0,j_0}$'s, we apply the BH procedure in \eqref{equation::BH-threshold-l} to obtain a rejection set $\cR_{\rm BH}^{i_0,k_0}(\alphaBH)$, as summarized in Algorithm \ref{algorithm::sample-local-test}. {The Algorithm accommodates extreme missing patterns (\eg, no observations in an entire row or no matched indices in $\Omega_{j_0}$) by setting $\wh A_{i_0,j_1}=1$ when $\Itraini=\emptyset$, and by applying \eqref{method::A-estimate} with $\widehat{d}_{j_1,j_2}=1$ when $\Omega_{j_0}=\emptyset$.}
We conclude this section with the FDR control guarantee for Algorithm \ref{algorithm::sample-local-test} in Theorem \ref{thm:local-FDR-split}.

\begin{theorem}[FDR control for the local test with empirical $p$-values by splitting] 
\label{thm:local-FDR-split}
Under model \eqref{def::exch-model} and 
Assumption \ref{assumption::rm}, for any $\alphaBH \in (0,1)$, {conditional on $\bM$}, the output $\cR_{\rm BH}^{i_0,k_0}(\alphaBH)$ of Algorithm \ref{algorithm::sample-local-test} for testing \eqref{def::hypothesis-testing-local-sampleV} satisfies FDR $\leq \alphaBH$ for testing \eqref{def::hypothesis-testing-local-sampleV}.
\end{theorem}

\begin{spacing}{0.925}
\begin{algorithm}
\small
\caption{Local test for \eqref{def::hypothesis-testing-local-sampleV} with FDR control
}
\label{algorithm::sample-local-test}
\KwIn{Target row index $i_0$, sets $\Itesti(k_0)$, $\Icalibi$, $\Itraini$, and calibration size $r_0 \ge C/\alphaBH$.}
        Partition $\Icalibi$ into disjoint $|\Itesti(k_0)|$ subsets randomly, i.e., $\{\Icalibi(j_0): j_0\in \Itesti(k_0)\}$\;

       \For{$j_0\in \Itesti(k_0)$}{
        Find the fully observed rows in the block $\{A(i,j)\}$, for $i\in \Itraini$ and $j\in \Icalibi(j_0)\cup\{j_0\}$, denoted by $\Omega_{j_0}$\;
 
         Calculate $\wh A_{i_0,j}$ for $j \in\Icalibi(j_0)\cup\{j_0\}$ using \eqref{def::d_{j_1,j_2,j}}--\eqref{method::A-estimate} {(if $\Itraini=\emptyset$, set $\wh A_{i_0,j_1}=1$; if $\Omega_{j_0}=\emptyset$, use \eqref{method::A-estimate} with $\widehat{d}_{j_1,j_2}=1$)}\;

        Compute p-value for $j_0$ using \eqref{eqn-def::emp-p-value}\;    
       }
       
       Apply the BH procedure in \eqref{equation::BH-threshold-l} to $p$-values in $\Itesti(k_0)$ and obtain $\cR_{\rm BH}^{i_0,k_0}(\alphaBH)$\; 

    \KwOut{Rejection set $\cR_{\rm BH}^{i_0,k_0}(\alphaBH)$.} 
\end{algorithm} 
\end{spacing}
\vspace{-0.35cm}

\subsection{Step 2: aggregating local tests for FDR control}
\label{section::2-3}


Recall that we break down the original problem in two stages: from an aggregated test to local tests, one for each row; then, within each row, we further break down the local test by partitioning $\Itesti$ into disjoint subsets. We now aggregate the results of the local tests to devise an {\it aggregated rejection set} with valid FDR control for predicting all missing links. To this end, we leverage the e-BH procedure \citep{wang2022false, ren2024derandomised}.

{The e-BH procedure offers a notable advantage that its validity requires only that each individual e-value satisfies $\mathbb{E}(e_{i_0,j})\leq 1$ under the null hypotheses in \eqref{def::hypothesis-testing-local-sampleV}, without any assumptions on their dependence structure.}
The central idea of our aggregated test is to rescale the binary test decision functions from the local tests into valid e-values.
Using the leave-one-out analysis of the BH procedure \citep{ferreira2006}, we show that $\mathbb{E}\big[ \mathbbm{1}\{ j\in \cR_{\rm BH}^{i_0,k_0}(\alphaBH) \}/\{|\cR_{\rm BH}^{i_0,k_0}(\alphaBH)| \vee 1\}\big] \le \alphaBH/|\Itesti(k_0)| $, for each $j$.
Therefore, for $j \in \Itesti(k_0)$,
\begin{align}
    e_{i_0,j} := \frac{|\Itesti(k_0)| \cdot \mathbbm{1}\{ j\in \cR_{\rm BH}^{i_0,k_0}(\alphaBH) \}}{\{|\cR_{\rm BH}^{i_0,k_0}(\alphaBH)| \vee 1\}\cdot \alphaBH}
    \label{equation::compute-e-value}
\end{align}
is a valid e-value, namely, $\mathbb{E}(e_{i_0,j})\leq 1 $ under the null $H_{i_0,j}$ in \eqref{def::hypothesis-testing-local}.
Then, by separately applying the e-BH procedure to $\{e_{i_0,j}, j \in \Itesti(k_0)\}$, one could obtain a local test for each row $i_0$ with the FDR controlled at level $\alpha_{\rm BH}$.
But here, we would directly construct an aggregated test using all $\{e_{i,j}:(i,j) \in \Dtest\}$. To this end, we sort all e-values $\{e_{i,j}:(i,j) \in \Dtest\}$ in descending order: $e_{(1)} \geq \cdots \geq  e_{(|\Dtest|)}$. Then, for any given $\alphaeBH>0$, set
\begin{align}
    \wh k
    :=
    \max\left\{
        k:
        1\leq k\leq |\Dtest|,
        e_{(k)} \geq \frac{|\Dtest|}{\alphaeBH k}
    \right\}~
    \text{ or }~ \wh k = 0 \text{ if the set is empty},
    \label{eqation::compute-hat-k}
\end{align}
and reject all $H_{i,j}$'s with e-values $e_{i,j}\geq e_{(\wh k)}$.
The procedure fails to reject any hypotheses when $\wh k = 0$.
This rejection rule controls FDR of \eqref{def::hypothesis-testing} at the level $\alphaeBH$.

To mitigate variations arising from the random selection of $\Icalibi$ and $\Itraini$, we further adopt a derandomization procedure \citep{ren2024derandomised}.
This builds on the property that the average of multiple e-values remains a valid e-value \citep{vovk2021values}.
For each row $i_0$ and $j \in \Itesti(k_0)$, we repeat Algorithm \ref{algorithm::sample-local-test} for $m_{i_0}$ times and compute the e-values $ e_{i_0,j}^{(1)},\ldots, e_{i_0,j}^{(m_{i_0})}$. Then we derandomize the e-values as follows:
\begin{align}
    \bar{e}_{i_0,j}
    :=
    \frac1{m_{i_0}}
    \sum_{l=1}^{m_{i_0}}
    e_{i_0,j}^{(l)},
    ~~
    \text{where } j \in \Itesti(k_0).
    \label{equation::compute-e-value-avg}
\end{align}
Finally, we apply the e-BH procedure to all $\bar e_{i,j}$'s as in \eqref{eqation::compute-hat-k}, using them in the same manner as $e_{i,j}$ in the original algorithm described above. This yields our method {\tt clp}, summarized in Algorithm \ref{algorithm::BH-EBH-framework}, which enjoys finite-sample valid FDR control as shown in Theorem \ref{thm:emp-fdr-global}.

\begin{spacing}{0.925}
\begin{algorithm}[h!]
\small
\caption{{\tt clp} method via e-value derandomization}
\label{algorithm::BH-EBH-framework}
    
    \KwIn{ Data $\big\{\bA\circ \bM^c, \bM\big\}$, 
    $\{c_{i,j}\}_{(i,j) \in \Dtest}$, 
    calibration size $r_0 \ge C/\alphaBH$, the number of derandomizations $m_{i_0}$ for $i_0 \in [n]$,
    and $\alphaBH, \alphaeBH \in (0, 1)$.}
   
    \For{$i_0 \in [n]$ where $|\Itesti|>0$}{

       Randomly construct $\Icalibi$ and partition $\Itesti$ into $k = \lceil |\Itesti| / r_1 \rceil$ approximately equal subsets, denote by $\Itesti(k_0)$, where $|\Itesti(k_0)| \leq r_1 := \lfloor |\Icalibi|/r_0\rfloor$ for $k_0\in [k]$ \;
         \For{$k_0\in [k] $}{
        Select $\Itesti(k_0)$ as the target test set\;

        \For{$m\in [m_{i_0}] $}{
        Randomly construct $\Icalibi$ and $\Itraini$\;
         
        Run Algorithm \ref{algorithm::sample-local-test} 
        and compute the e-values $\{e_{i_0,j}^{(m)}:j \in \Itesti(k_0)\}$ via \eqref{equation::compute-e-value}\;
        }
     Compute the e-values $\big\{\bar{e}_{i_0,j}:j \in \Itesti(k_0)\big\}$ using \eqref{equation::compute-e-value-avg}\;
     }  
    }
    
    Record e-values $\{\bar{e}_{i,j}:(i,j) \in \Dtest\}$ and compute $\wh k$ using \eqref{eqation::compute-hat-k}\;
    
    \KwOut{The rejection set $\cR_{\rm eBH}:=\left\{(i,j) \in \Dtest:\bar{e}_{i,j} \geq |\Dtest| / (\alphaeBH \wh k )\right\}$.}
\end{algorithm}
\end{spacing}
\vspace{-0.05cm}

\begin{theorem}[FDR control of {\tt clp} for testing \eqref{def::hypothesis-testing}]
\label{thm:emp-fdr-global} 
    Under the conditions of Theorem \ref{thm:local-FDR-split}, given any $\alphaeBH\in (0,1)$ and for any choice of $\alphaBH \in (0,1)$, {conditional on $\bM$,} the rejection set produced by Algorithm \ref{algorithm::BH-EBH-framework} satisfies FDR $\leq \alphaeBH$.
\end{theorem}

We make two key remarks.
First, our method {\tt clp} accommodates arbitrary missing mechanism, provided Assumption \ref{assumption::rm} hold. Examples include: (1) random but non-uniform missingness \citep{gui2023conformalized}, and (2) structured missingness, such as block or staggered patterns commonly encountered in panel data \citep{athey2021matrix,xiong2023large}.
Second, as shown in Theorem \ref{thm:emp-fdr-global}, {\tt clp} guarantees valid FDR control for any choice of $\alphaBH$, while \cite{ren2024derandomised} recommend setting $\alphaBH=\alphaeBH/2$ for improved powers.

We note that our method may exhibit some conservativeness in practice.
By \eqref{equation::compute-e-value}, the maximum possible value of $e_{i_0,j}$ for $j \in \Itesti(k_0)$ is $|\Itesti(k_0)|/\alphaBH$. Thus, when $|\Itesti(k_0)|$ is small due to extensive data-splitting, the resulting e-values may be lowered, limiting the number of rejections under the e-BH procedure in \eqref{eqation::compute-hat-k}.
To alleviate this effect, we introduce an inflation factor to adjust each e-value $e_{i,j}$ and increase the likelihood of rejection. Numerical studies in Section \ref{section::numerical-studies} explore distinct choices of this inflation factor and suggest that using a factor of $1/\alphaBH$ effectively enhances  power while maintaining well-controlled FDR.
A more rigorous theoretical characterization of this practically effective factor remains an intriguing yet challenging direction for future work.


Computing the dissimilarity measure in \eqref{def::d_{j_1,j_2}} is computationally intensive due to the evaluation of all terms $\widehat d_{j_1,j_2,j}$ in \eqref{def::d_{j_1,j_2,j}}. 
The overall computational complexity of {\tt clp} is $O\big(m|\Itraini|^2|\Icalibi|\big)$, where $m$ is the number of missing entries, assuming $|\Itraini|$ (and $|\Icalibi|$) are of the same order across all $i_0 \in [n]$. Specifically, computing \eqref{def::d_{j_1,j_2,j}} and \eqref{def::d_{j_1,j_2}} incurs a cost of $O\big(|\Itraini|\big)$, while the aggregation step in \eqref{method::A-estimate} adds additional $O\big(|\Icalibi|)$. In practice, our method can be accelerated by reducing the sizes of $\Itraini$ and $\Icalibi$; for instance, setting $\Itraini\asymp\Icalibi=O(1)$ reduces the overall complexity to $O(m)$, though at the potential cost of reduced accuracy in estimating dissimilarity and, consequently, lower power.

\subsection{Extensions and comparisons}
While our proposed framework is detailed for directed networks, it naturally extends to bipartite and undirected networks with minimal modifications. The adaptation to bipartite networks is straightforward, owing to the their exchangeability. That is, we can use more rows to construct $\wh A_{i_0,j}$ and the associated $p$-values. Extending to undirected networks requires extra care due to the symmetry constraints $A_{i,j}=A_{j,i}$ and $M_{i,j}=M_{j,i}$, but the solution is remarkably simple.
We first apply our Algorithms \ref{algorithm::sample-local-test} and \ref{algorithm::BH-EBH-framework} up to Line 9 of Algorithm \ref{algorithm::BH-EBH-framework}, where the $\bar e_{i,j}$ values are obtained. Thereafter, we apply the remainder of Algorithm \ref{algorithm::BH-EBH-framework}, the e-BH procedure, only to the upper triangular part, i.e., $\{\bar e_{i,j}: (i,j)\in \Dtest \text{ and }i<j\}$, instead of the full set $\{\bar e_{i,j}: (i,j)\in \Dtest\}$. Formal algorithmic descriptions (Algorithms \ref{algorithm::sample-local-test-und} and \ref{algorithm::sample-local-test-bipartite}) and theoretical justifications for both extensions are provided in Section \ref{supp-section::bipartitie-and-undirected} of the Supplementary Material.
 
We conclude this section by comparing our method with existing works.
\citet{gui2023conformalized} considers arbitrary $\bA$ under a missing-at-random mechanism but requires knowledge of, or accurate estimates for, the odds ratios introduced by missing probabilities.
In contrast, we allow arbitrary missingness patterns $\bM$ that are independent of $\bA$, and leverage the latent graphon structure in $\bA$ to reinstate the exchangeability. This enables our method to accommodate a  broader class of missing mechanisms.
Overall, the two approaches rely on different assumptions and do not subsume each other. {We also note that our paper considers random matrices $\bA$, whereas in related work such as \cite{gui2023conformalized,liang2024structured}, the matrix to be tested is treated as fixed}. More recently, \citet{marandon2024conformal, blanchard2024fdr} proposed a novel FDR control approach that casts link prediction as a classification problem. 
While innovative, their formulation is inherently limited to unweighted networks and assumes missingness rates take only two distinct values, depending on whether an entry is $0$ or $1$. In contrast, our method applies to both weighted and unweighted networks and allows fully heterogeneous, unknown missingness rates, offering greater flexibility in practice. {It is also possible to leverage general p-to-e calibrators \citep{Shafer2011,vovk2021values} to construct the e-values, but as reported in literature, this approach tends to be rather conservative.
We leave further studies along this line to future research.}
 
\section{Simulations}
\label{section::numerical-studies}

\subsection{Simulation 1: Tuning our algorithm}
\label{section::numerical-studies-1}

We generate networks following the distribution $A_{i,j} = f(\xi_i,\xi_j) + \epsilon_{i,j}$, where $\epsilon_{i,j} \sim \mathrm{Uniform}[-0.1,0.1]$.
Three graphons are considered to cover a wide range of scenarios: 

\noindent
(i) Setting 1 (low-rank and smooth): $f_1(\xi_i,\xi_j) = \xi_i^3 + 2\xi_j^3$; 
    
\noindent
(ii) Setting 2 (high-rank): $f_2(\xi_i,\xi_j) = \max(\xi_i, \xi_j)^{\frac{2}{3}}\cos[0.1/\{(2 \xi_i-1/2)^3 + (\xi_j-1/2)^3 + 0.01\}]$;
    
\noindent
(iii) Setting 3 (non-smooth): $f_3(\xi_i,\xi_j) =  (3 \xi_i^2 + \xi_j^2) \cdot \cos\{{(2 \xi_i^4 + \xi_j^4)}^{-1}\}$. 

We fix the network size at $n=200$ and set the proportion of missing links to $0.2$, resulting in an average of $3,980$ missing links. Specifically, we generate $M_{i,j}\sim$ Bernoulli$(q_{i,j})$ and consider heterogeneous missing probabilities, where $q_{i,j}$'s are independently generated from $\mathrm{Uniform}[0,0.4]$.
We set the null hypotheses to be true, namely $c_{i,j} = A_{i,j}$ in \eqref{def::hypothesis-testing}, for $70$\% of entries.
For the remaining $30\%$, we set $c_{i,j} = A_{i,j} - 1.5$.
For each row $i_0$, all observed entries are randomly split into $\Itraini \cup \Icalibi$ with a ratio of $|\Itraini|:|\Icalibi|=2:3$, and we set $\alphaBH = \alphaeBH/2$ following \citet{ren2024derandomised}.
In the BH procedure, we choose $r_0=25$, as defined in the paragraph following \eqref{def::hypothesis-testing-local-sampleV}. We use the Gaussian kernel
$K(x) = \exp\big(-x^2/2\big)/\sqrt{2\pi}$ in \eqref{method::A-estimate} and derandomize local tests using $20$ repetitions in \eqref{equation::compute-e-value-avg}.

\begin{figure}[h]
    \centering
    \includegraphics[width=0.8\textwidth]{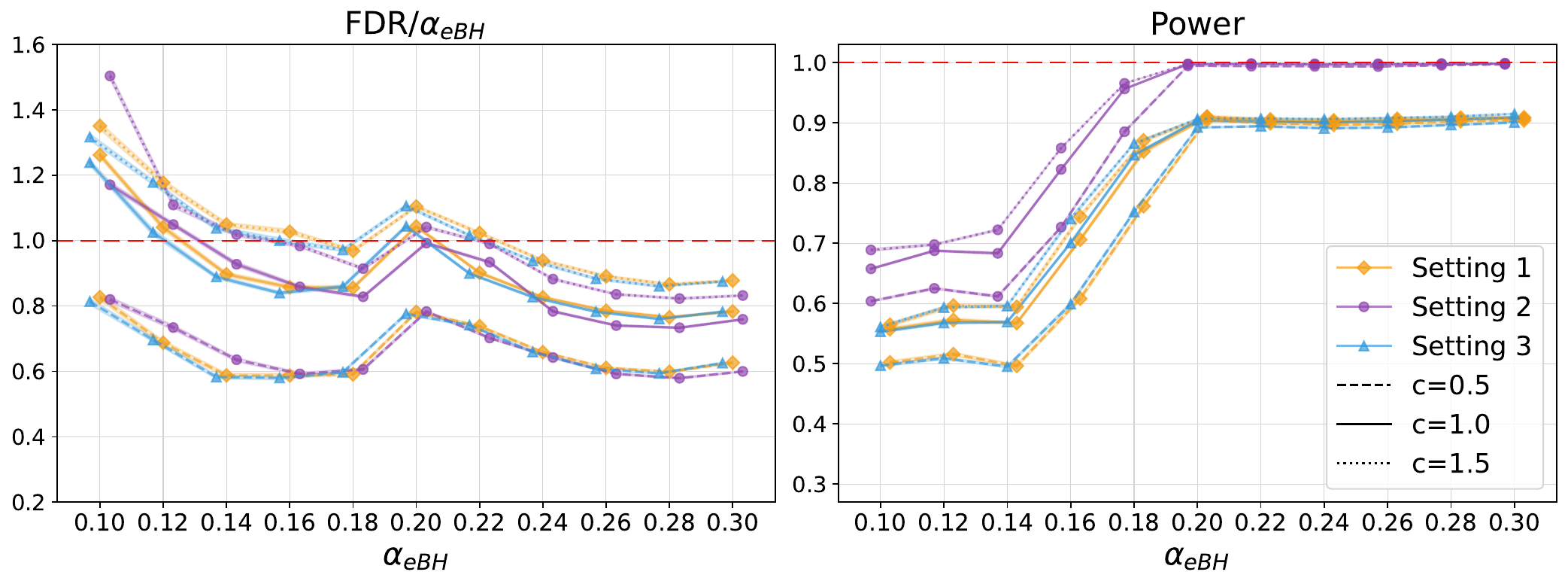}
    \caption{Empirical ratio of FDR and $\alphaeBH$, and empirical power under varying $\alphaBH$ and $\alphaeBH$. }
    \label{fig::infl-c}
\end{figure}

Recall that in \eqref{equation::compute-e-value}, we inflate the e-values by a factor of $c/\alphaBH$.
Here, we examine the impact of $c\in\{0.5, 1, 1.5\}$.
Performance is evaluated using the FDR, approximated by the averaged FDP, 
and power, both assessed over $100$ experiments.
We vary $\alphaeBH$ from $0.1$ to $0.3$. The results in Figure \ref{fig::infl-c} show that our method consistently controls the empirical FDR across significance levels when $c=0.5$, and in most settings when $c=1$. It also achieves desirable power results as $\alphaeBH$ increases. Overall, we recommend $c=1$ as it strike a satisfactory balance between FDR control and  decent power.

\subsection{Simulation 2: Comparison with benchmark methods}
\label{section::numerical-studies-2}

We compare our method ({\tt clp}) with conformal matrix completion ({\tt cmc}, \cite{gui2023conformalized}) and 
the link prediction method by \cite{marandon2024conformal} ({\tt dss}) via two sub-simulations for weighted and unweighted networks, respectively.
While {\tt cmc} applies to both settings, {\tt dss} is, by design, primarily suited for unweighted networks.

For the first sub-simulation (weighted links), we adopt Settings 1--3 from Section \ref{section::numerical-studies-1}, fixing $\alphaBH=0.1$ and $\alphaeBH=0.2$.
The threshold $c_{ij}$ is set as a specific quantile of the observed link weights, with the quantile level $\kappa$ varying from $0.1$ to $0.5$. As noted earlier, we inflate the e-values in \eqref{equation::compute-e-value} by a factor of $1/\alpha_{\text{BH}}$. 
Note that {\tt cmc} only provides confidence intervals with guaranteed coverage, rather than $p$-values. Its underlying idea is to approximate the population quantiles of absolute prediction errors by the empirical quantiles of calibrated absolute errors, a procedure valid only for two-sided problems.
To adapt it to our one-sided multiple testing problem \eqref{def::hypothesis-testing}, we modify it to produce comparable $p$-values: we omit the absolute value when computing errors and construct the empirical distribution function $\wh{F}_{\text{eCDF}}(\cdot)$ based on the calibrated errors $\wh{A}_{ij}-A_{ij}$. We then compute $p$-values for {\tt cmc} as $1-\wh{F}_{\text{eCDF}}(\wh{A}_{ij}-c_{ij}), (i,j)\in\Dtest$, and apply the BH procedure. Note also that {\tt cmc} requires a hypothesized rank as input. In practice, results of {\tt cmc} are robust to the hypothesized rank, as also noted by \citet{gui2023conformalized}. Accordingly, we fix the rank at $r=2$ for {\tt cmc} across all settings for demonstration.

\begin{figure}[h]
    \centering
    \includegraphics[width=0.8\textwidth]{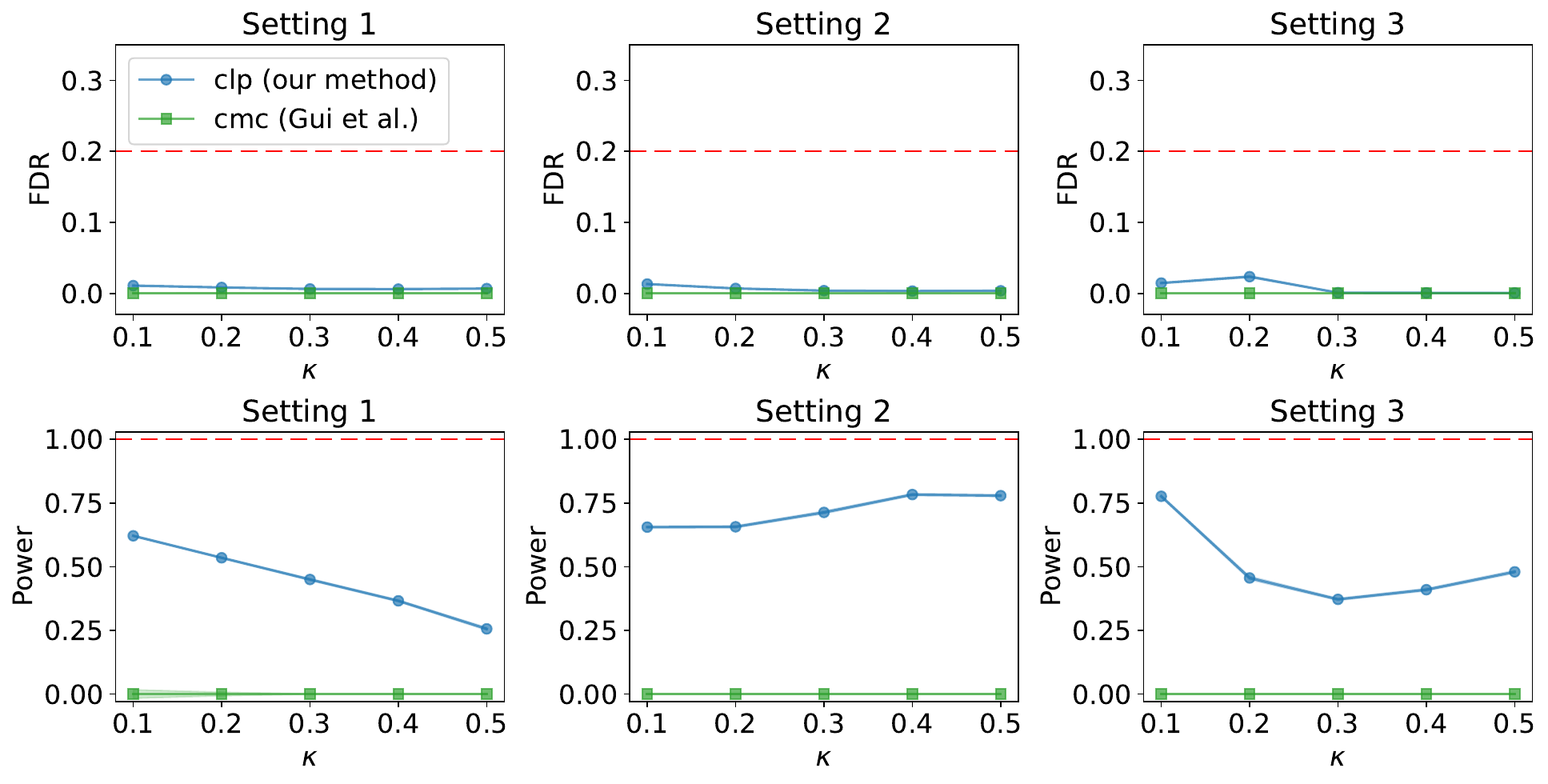}
    \caption{Comparisons of empirical FDRs and powers across different $c_{ij}$'s under Settings 1--3, with $\alphaBH=0.1$ and $\alphaeBH=0.2$. 
    }
    \label{fig::weight-unif-c}
\end{figure}

As shown in Figure \ref{fig::weight-unif-c}, while both {\tt clp} and {\tt cmc} control the FDR, {\tt clp} consistently achieves higher power across all settings. 
We note that {\tt cmc} is sensitive to the choice of $c_{i,j}$ values. In Section \ref{supp::additional-numerical-studies} of the Supplementary Material, we adopt the approach in Section \ref{section::numerical-studies-1} to select $c_{i,j}$ based on signal strength. Under such configuration, {\tt cmc} fails to control the FDR, whereas {\tt clp} continues to perform well.

To compare with {\tt dss}, we set up a second sub-simulation for unweighted networks, using a different network-generating process from that used for the weighted case. Specifically, we generate $A_{i,j} = \mathbbm{1}\{(\xi_i+\xi_j)/2 + \epsilon_{i,j}>t\}$, where $\epsilon_{i,j} \sim \mathrm{Uniform}[-0.25,0.25]$.
We set $c_{i,j}=0.5$ in \eqref{def::hypothesis-testing} and vary $t \in \{0,0.05,0.1,0.15,0.2\}$ to control the number of null and alternative hypotheses. 
We consider heterogeneous missingness by drawing $q_{i,j}\sim \mathrm{Uniform}[0,0.2]$ independently for each $(i,j)$, yielding an average proportion of missing links of approximately $10\%$. We fix $\alphaBH=0.2$ and $\alphaeBH=0.2$, and use $r_0=50$ to improve estimation of $A_{i,j}$ for unweighted networks.

\begin{figure}[h]
    \centering
    \includegraphics[width=0.65\textwidth]{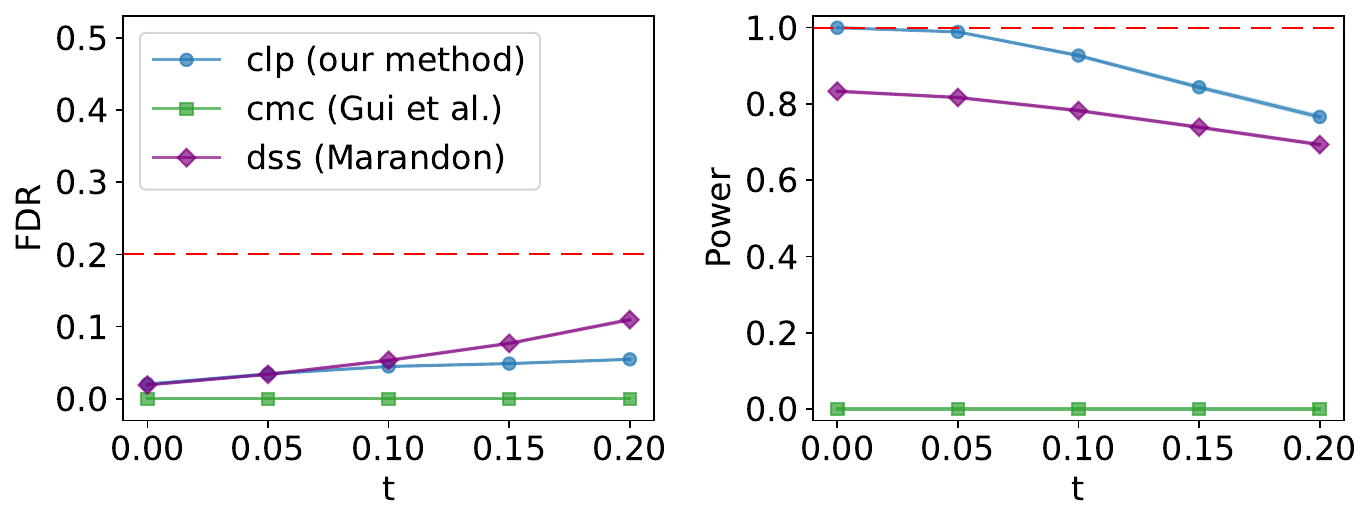}
    \caption{Comparisons of empirical FDRs and powers for unweighted networks with $\alphaBH=0.2$ and $\alphaeBH=0.2$. 
    }
    \label{fig::unweight-unif-c}
\end{figure}

As displayed in Figure \ref{fig::unweight-unif-c}, all three methods control the FDR across different values of $t$.
The conservativeness of {\tt cmc} aligns with our observation for weighted networks.
While the FDR of {\tt dss} exceeds that of {\tt clp}, its relative low power can be attributed to the heterogeneous missing probabilities. In general, both {\tt clp} and {\tt dss} outperform {\tt cmc} in the unweighted network setting, and our method demonstrates strong performance across both weighted and unweighted networks.
 

\section{Data example: bilateral trade flows}
\label{section::real-data}

We consider the bilateral trade flow data from \citet{helpman2008estimating}, which records trade volumes between $158$ countries in 1989. The dataset forms a directed network, where both indices $i$ and $j$ represent countries acting as exporters and importers, respectively. The outcome ${A_{i,j}}$ denotes the log-transformed trade volume from country $i$ to country $j$, with a mean of $3.82$, standard deviation of $4.69$, and a range from $0$ to $18.39$. Here,  $A_{i,j}=0$ indicates no recorded trade flow. 
We randomly mask $10\%$ of the links, corresponding $2,496$ trade entries, for evaluating the prediction performance of different methods.
For these held-out entries, we set $c_{i,j} = 0.2$ in the random hypotheses \eqref{def::hypothesis-testing}, resulting in approximately equal proportion of null and alternative hypotheses.

\begin{figure}[h]
    \centering
    \includegraphics[width=0.8\textwidth]{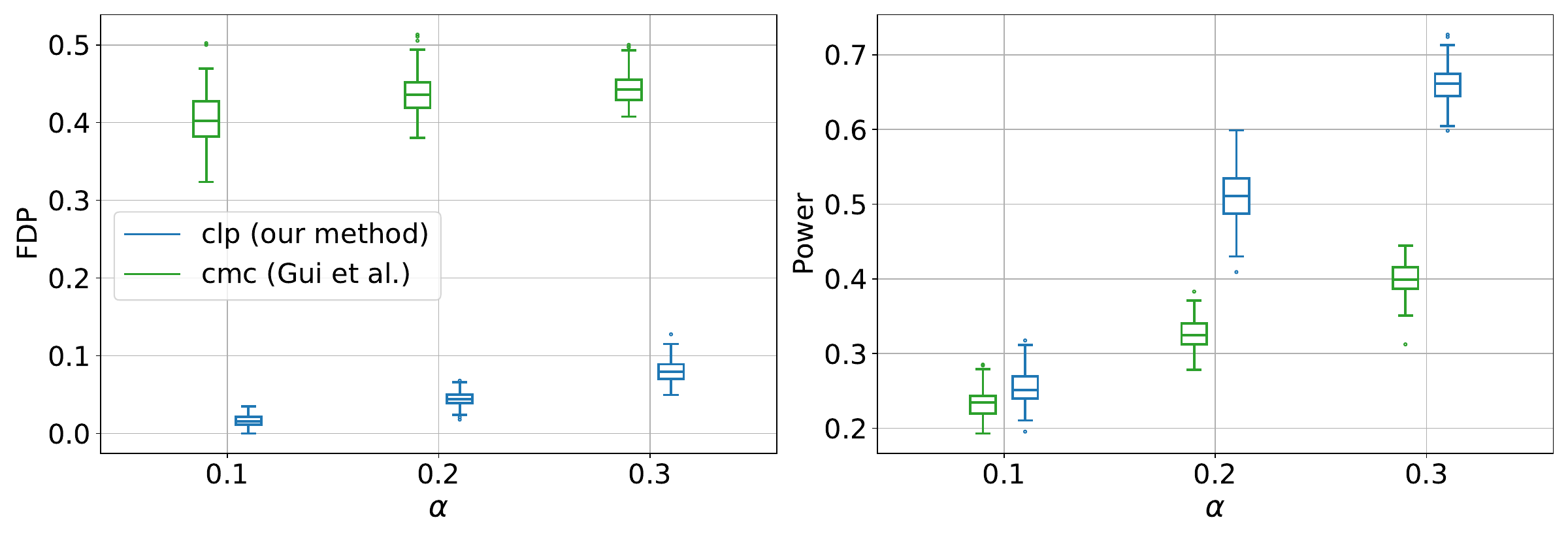}
    \caption{Box-plots of FDPs and powers over $100$ independent replications for predicting missing links in the international trade dataset.}
    \label{fig::trade}
\end{figure}

We compare our method only with {\tt cmc}, as the method {\tt dss} is not applicable because it is classification-based and limited to binary or categorical links. Following our simulation setup, we implement {\tt clp} with $|\Itraini|:|\Icalibi|=2:3$ for each $i_0 \in [n]$, set $r_0=25$, and use $\alphaBH=\alphaeBH/2$. 
We employ a Gaussian kernel for $K(\cdot)$ in \eqref{method::A-estimate-oracle} and perform $20$ derandomizations in \eqref{equation::compute-e-value-avg}. The nominal FDR level $\alpha$ is varied over $\{0.1, 0.2, 0.3\}$.


The left panel of Figure \ref{fig::trade} shows that {\tt clp} controls FDR well across all the nominal FDR levels, whereas {\tt cmc} substantially inflates the FDP. The right panel demonstrates that our method is also more powerful than {\tt cmc}. As $\alpha$ increases, the empirical power of {\tt clp} improves, and its advantage over {\tt cmc} becomes more pronounced.

\section{Concluding remarks}

In this article, we develop a general framework for FDR control in multiple network link prediction, leveraging e-values and insights from conformal inference.
We propose data-driven methods and a multiple-splitting strategy to reinstate the exchangeability required for computing nonconformity scores under complex network edge dependencies and heterogeneous missingness. With supporting theories, we establish performance guarantees for multiple link prediction without requiring knowledge of the underlying network distribution. 

Experiments on both simulated and real data suggest that a $1/\alphaBH$ inflation factor performs well for {\tt clp}. 
Setting a conservative threshold for e-values in the e-BH procedure may be inefficient in practice. \cite{blier2024improved} proposed some preliminary strategies for boosting e-values in the e-BH procedure under certain distributional assumptions while maintaining FDR control.
It is of interest to investigate whether an optimal inflation criterion of e-values exists in our context, and if so, how it can be effectively designed. 

{We note that block-wise schemes may be designed to offer computational savings. However, such approaches are challenging in our setting, as the required block-level exchangeability is fragile under random missingness in $\bM$. Missing entries can break the symmetry needed for valid permutation or aggregation, making it difficult to construct blocks that both satisfy exchangeability and remain robust to heterogeneous missingness patterns. Moreover, forming blocks using only observed entries limits their applicability for testing missingness itself. We leave a rigorous treatment of block-wise aggregation under missingness to future work.}





\section*{Code}
The code for reproducibility is available at: \url{https://github.com/code-cloud9/CLP}.

\section*{Acknowledgement}
Dong Xia was supported by Hong Kong RGC grant GRF 16303224.
Yuan Zhang was supported by United States NSF grant DMS-2311109.
Wen Zhou was supported by United States NIH grants R01GM157600 and R01GM144961.

\bibliographystyle{apalike}
\bibliography{paper-ref}


\newpage
\appendix

\begin{center}{\bf \Large  Supplementary Material for\\``Conformal network link prediction with false discovery rate control under unstructured missingness"}

\end{center}

\noindent
This Supplementary Material contains proofs of Theorems \ref{thm::FDR-rowwise}--\ref{thm:emp-fdr-global} and Propositions \ref{prop::withinrow-exch}--\ref{prop:emp-cond-ind} in Section A. Additional results of numerical studies are provided in Section \ref{supp::additional-numerical-studies} for simulations and Section \ref{supp::additional-real-data-analysis} for real data application. The extensions of our proposed method {\tt clp} to undirected and bipartite networks are given in Section \ref{supp-section::bipartitie-and-undirected}. {Additional results for Bernoulli missing patterns are given in Section \ref{supp::Bern-M}.} Unless stated otherwise, all the notation follows the same definitions as in the main body of the paper.

\renewcommand\thesection{\Alph{section}}
\setcounter{section}{1}

\renewcommand\theequation{B.\arabic{equation}}
\renewcommand\thelemma{B.\arabic{lemma}}
\setcounter{equation}{0}
\setcounter{lemma}{0}
\setcounter{figure}{0}
\renewcommand\thefigure{B.\arabic{figure}}

\section{Additional numerical studies}
\label{supp::additional-numerical-studies}

We explore how signal strength affect the performance of {\tt clp}, {\tt cmc} and {\tt dss}. 
For weighted network setting, we apply the graphons in Section \ref{section::numerical-studies-1}, setting $c_{i,j} = A_{i,j} - \delta$ for $30$\% of the hypotheses and set $c_{i,j} = A_{i,j}$ for the rest as the null hypothesis. Here, $\delta$ controls signal strength.

\begin{figure}[h]
    \centering
    \includegraphics[width=0.95\textwidth]{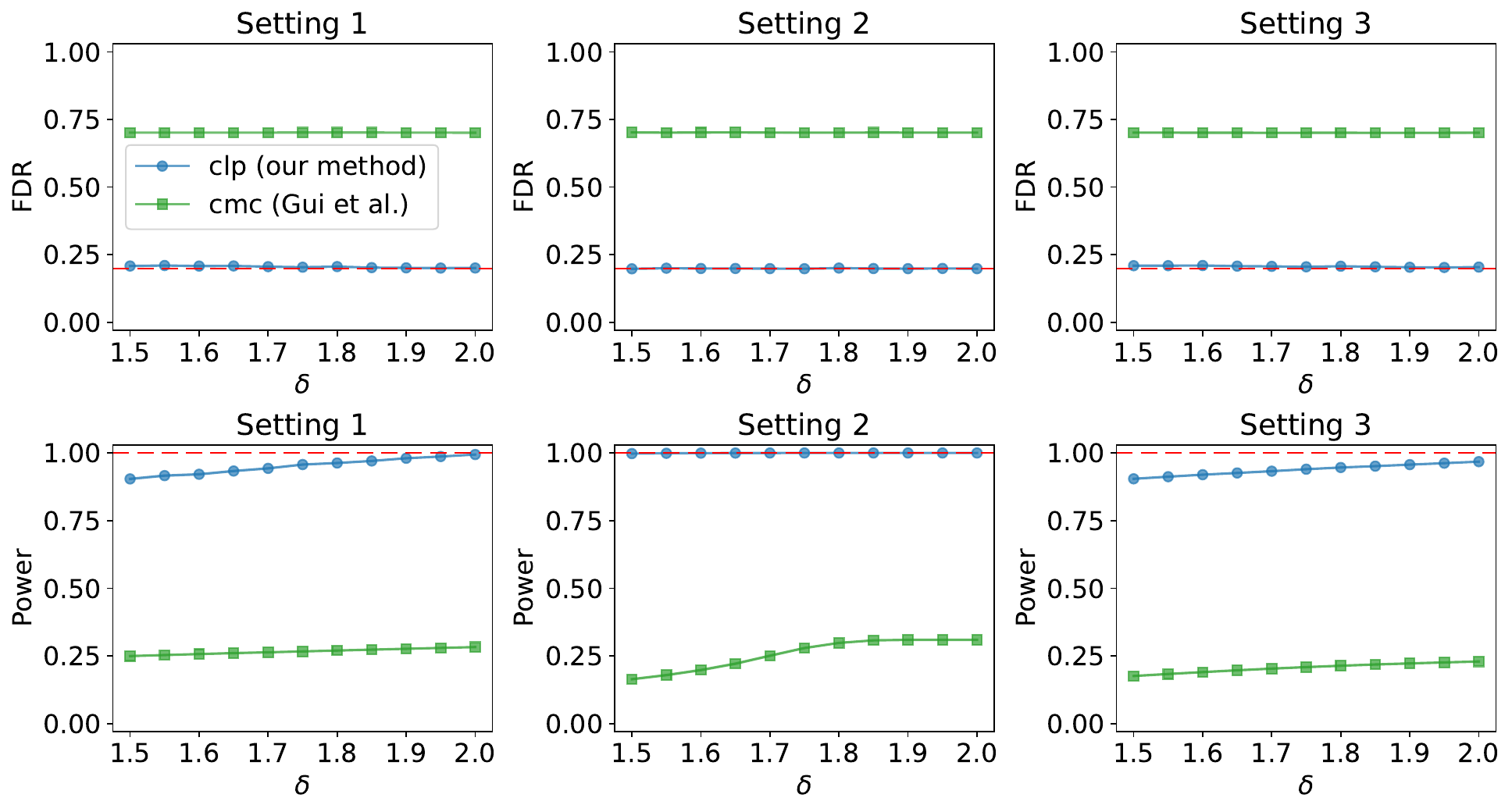}
    \caption{Comparison results across different signal strengths through empirical FDR and power under Settings 1--3 with $\alphaBH=0.1$ and $\alphaeBH=0.2$. 
    }
    \label{fig::supp-weighted-signal}
\end{figure}

Figure \ref{fig::supp-weighted-signal} shows the empirical FDR and power across different values of signal strength $\delta$.
As can be seen, {\tt cmc} fails to control the FDR across all settings.
Comparing the results in Figure \ref{fig::supp-weighted-signal} and Figure \ref{fig::weight-unif-c}, we see that the performance of {\tt cmc} is sensitive to the choice of $c_{ij}$'s. 
In contrast, {\tt clp} consistently achieves desirable FDR control and its power increases as the signal strength becomes larger across all settings.

Since the signal strength is not straightforward to define in the unweighted setting, we slightly modify our testing target to illustrate how signal strength affects the performance of different methods.
Specifically, we generate $p^*_{i,j} = f(\xi_i,\xi_j) + \epsilon_{i,j}$ with the graphon functions in Settings 1--3 and rescale them by $p_{i,j}=(p^*_{i,j}-\min_{i,j}p^*_{i,j})/(\max_{i,j}p^*_{i,j}-\min_{i,j}p^*_{i,j})$ to $[0,1]$-valued probabilities.
Then we independently generate $A_{i,j} \sim \text{Bernoulli}(p_{i,j})$. Different from the previous settings, here we consider heterogeneous missing probabilities where $q_{i,j}$'s are independently generated from $\mathrm{Uniform}[0,0.4]$. 
Instead of testing whether there is a true link or not, we test $H_{i,j}: p_{i,j} \leq c_{i,j}$ for the random $p_{i,j}$'s, and set $c_{i,j} = p_{i,j} - \delta$ for 30\% of the hypotheses regarding the missing entries. Otherwise, we set $c_{i,j} = p_{i,j}$ for the null hypotheses. 

\begin{figure}[h]
    \centering
    \includegraphics[width=0.95\textwidth]{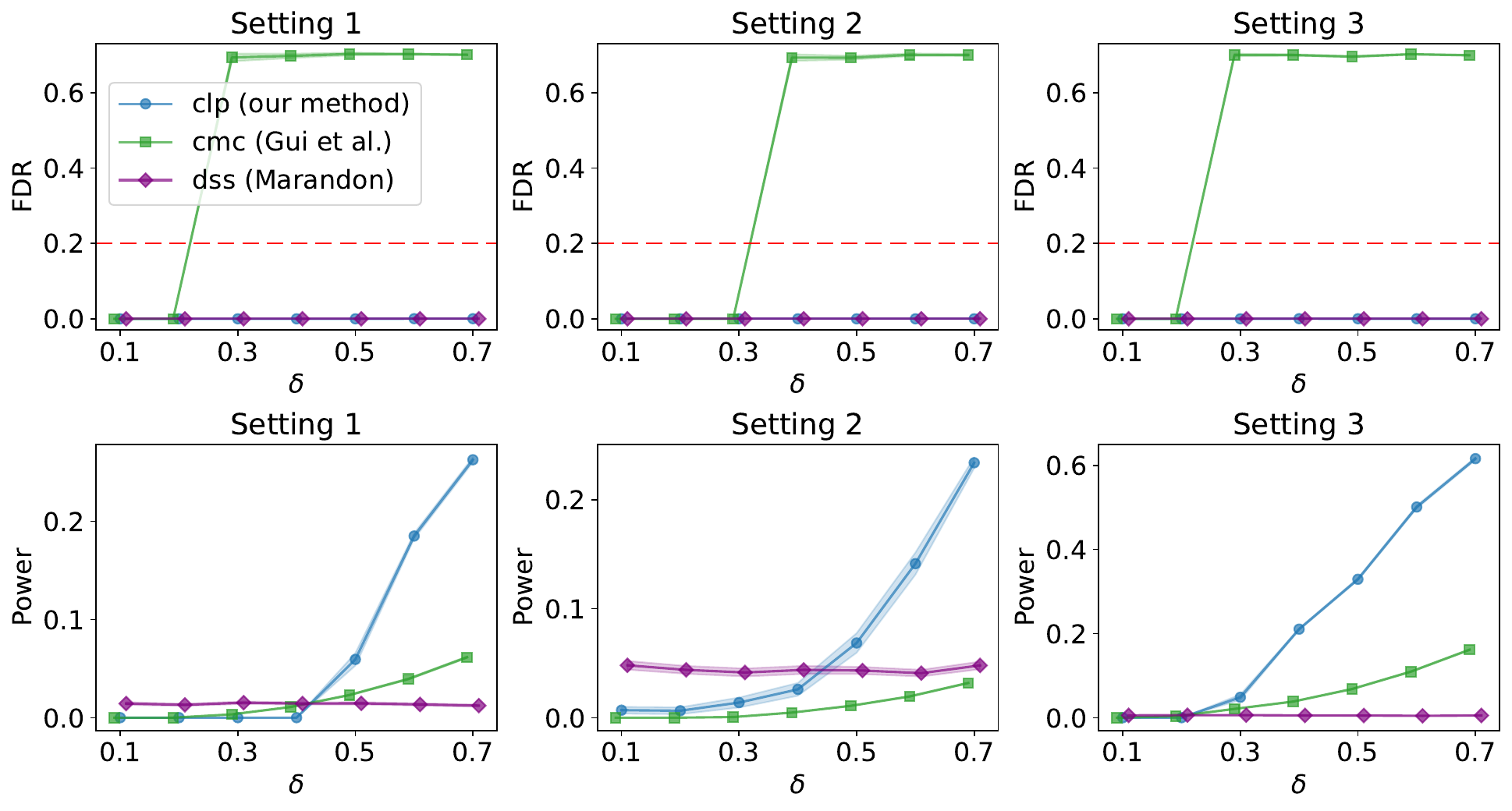}
    \caption{Comparison results through empirical FDR and power under modified Settings 1--3 for unweighted networks with $\alphaBH=0.1$ and $\alphaeBH=0.2$. 
    }
    \label{fig::binary-comparison}
\end{figure}

As shown in Figure \ref{fig::binary-comparison}, {\tt clp} and {\tt dss} control FDR throughout the three settings, while {\tt cmc} fails.
Additionally, the relative low power of {\tt dss} compared to {\tt clp} can be attributed to the heterogeneous missing probabilities. \citet{marandon2024conformal} assumes that the missingness rates $q_{i,j}$'s take only two distinct values, depending on whether the links are $0$ or $1$. The target of testing random $p_{i,j}$'s instead of $A_{i,j}$'s can also contribute to {\tt dss} making fewer rejections.

{
\subsection{Comparison with {\tt dss} under weighted networks}
\label{supp::Comparison-with-Marandon-weighted}


Note that the procedure in \cite{marandon2024conformal} ({\tt dss}) cannot be easily extended to weighted network settings with heterogeneous null values $c_{i,j}$, since their $D_{\rm train}^0$ (the set of observed non-existent edges) and $D_{\rm train}^1$ (the set of observed true edges) are not well-defined when $c_{i,j}$ differs across missing entries.
Furthermore, in Settings $1$--$3$ introduced in Section \ref{section::numerical-studies-1} of the main paper, where the $c_{i,j}$'s are uniform, one way to apply the procedure in \citet{marandon2024conformal} to test 
$
    H_{i,j}: A_{i,j} \leq c_{i,j}~ 
    \text{for } (i,j) \in {\cal D}_{\rm test},
$
is as follows: during testing, we first transform the observed weighted network into an unweighted network using the threshold $c_{i,j}$ (set to the quantile of the observed links in Settings $1$--$3$). We then apply the procedure in \citet{marandon2024conformal} to this transformed unweighted network.
\begin{figure}[h]
    \centering
    \includegraphics[width=0.9\textwidth]{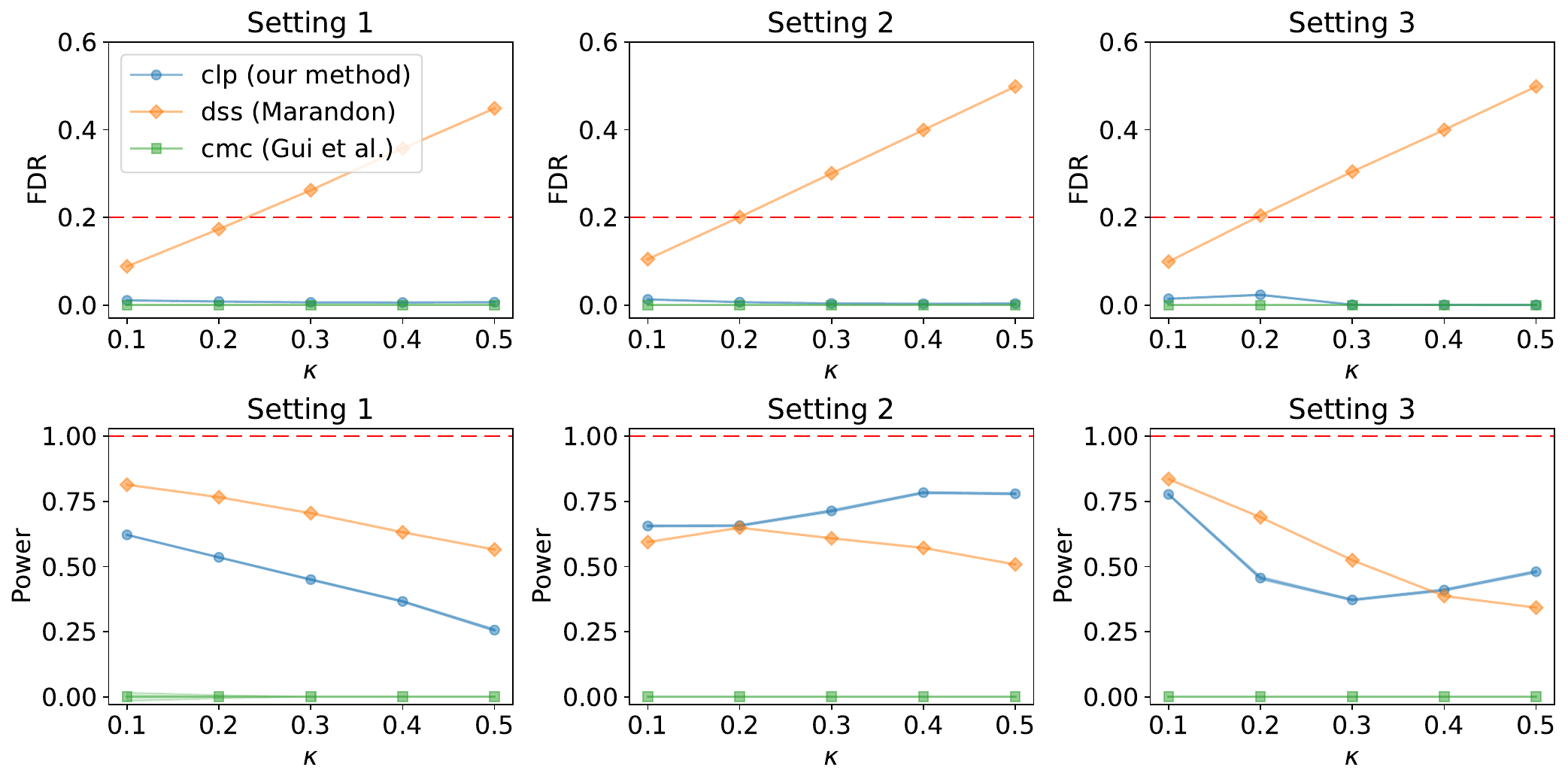}
\caption{Comparisons of empirical FDRs and powers of different methods under Settings 1--3.}   
\label{fig::dss-binary}
\end{figure}
The results are shown in Figure \ref{fig::dss-binary}. We observe that the procedure in \citet{marandon2024conformal} fails to control the FDR. This can be attributed to the heterogeneous missing probabilities in Settings $1$--$3$, which are not addressed by their method, as it assumes homogeneous missing probabilities. 
%

\subsection{Sensitivity of {\tt cmc} to the choice of hypothesized rank}

For the {\tt cmc} method, we conducted additional experiments using a larger hypothesized rank, namely $r=10$, under Settings 1--3 introduced in Section \ref{section::numerical-studies-1} of the main paper. As shown in Figure \ref{fig::cmc-r10}, the performance of {\tt cmc} with $r=10$ is similar to that with $r=2$ in these settings, in terms of both empirical FDR and power. As noted by \citet{gui2023conformalized}, in practice, results of {\tt cmc} are robust to the hypothesized rank. Therefore, the performance of {\tt cmc} shown in Figure \ref{fig::cmc-r10} can be attributed to the inherent robustness of the {\tt cmc} method and that in Settings $ 1 $--$ 3 $, increasing $r$ to $10$ may not alter the ability to recover the core low-dimensional signal in the network. 

\begin{figure}[h]
    \centering
    \includegraphics[width=0.9\textwidth]{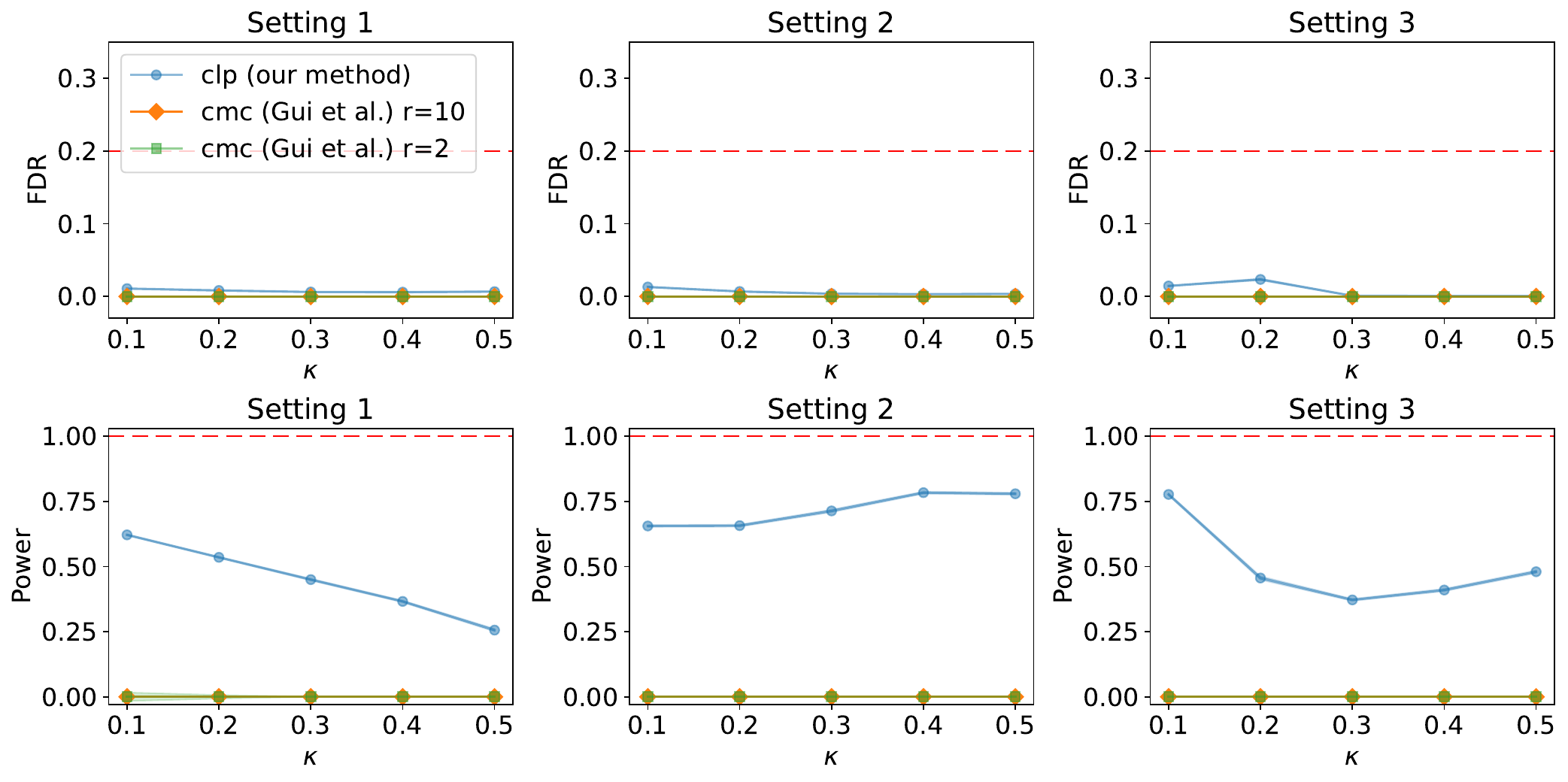}
\caption{Comparisons of empirical FDRs and powers for the {\tt cmc} method with $r=2$ and $r=10$ under Settings 1--3.}   
\label{fig::cmc-r10}
\end{figure}

\subsection{Weighted and unweighted estimators in {\tt clp}}
\label{supp::weighted-unweighted}


Recall that in Section \ref{sec23} of the main paper, we construct a ``fine-grained weighted estimator" of $\bA$ through the weights
$
    w(j_1,j_2),
$
which are determined by a dissimilarity measure $d(j_1,j_2)$. The purpose of this construction is to preserve the relevant exchangeability structure while improving the estimation accuracy of $\widehat{\bA}$. A natural question is that whether one could instead use an unweighted estimator, that is, an estimator with constant weights, and thereby pursue a trade-off between estimator simplicity and the estimation accuracy.
The answer is yes. In fact, the unweighted estimator can be viewed as a special case of our general construction. Specifically, if we take
$
    d(j_1,j_2)\equiv 1,
$
then the resulting weight becomes
$
    w(j_1,j_2)=\frac{1}{|\Itraini|},
$
which is constant over $j_2\in \Itraini$. Hence, the corresponding estimator is unweighted. Moreover, the exchangeability argument underlying our {\tt clp} procedure continues to apply in this case. Therefore, from a validity perspective, the unweighted estimator is also admissible.

However, using constant weights may substantially degrade the estimation accuracy of $\wh \bA$, which may consequently reduce power. To illustrate this trade-off, we conduct additional simulations comparing the weighted estimator and the unweighted estimator. We consider both a weighted-network setting and an unweighted-network setting. For the weighted-network setting, we use Setting 3 in Section \ref{section::numerical-studies-1} of the main paper. For the unweighted-network setting, we use the data-generating mechanism introduced in Section \ref{section::numerical-studies-2}. In both cases, we apply {\tt clp} with the same implementation as in the main paper, including the $1/\alpha_{\rm BH}$ inflation factor for the e-values. Figure \ref{fig::we-unwe-comparison} reports the empirical FDR and power for the two estimators.
We present our original {\tt clp} results using the ``fine-grained weighted estimator'', shown in blue and labeled as the ``weighted estimator''. For comparison, the results using the ``unweighted estimator" sets $d\left(j_1, j_2\right) \equiv 1$, as discussed above. Its performance is shown in orange.
Under the weighted-network setting, the weighted estimator and the unweighted estimator perform similarly. Under the unweighted-network setting, the weighted estimator yields better power than the unweighted estimator, while still maintaining FDR control. These results suggest that although the unweighted estimator is valid and conceptually simpler, the fine-grained weighted estimator can have better power performance in practice. 

\begin{figure}[h]
    \centering
    \includegraphics[width=\textwidth]{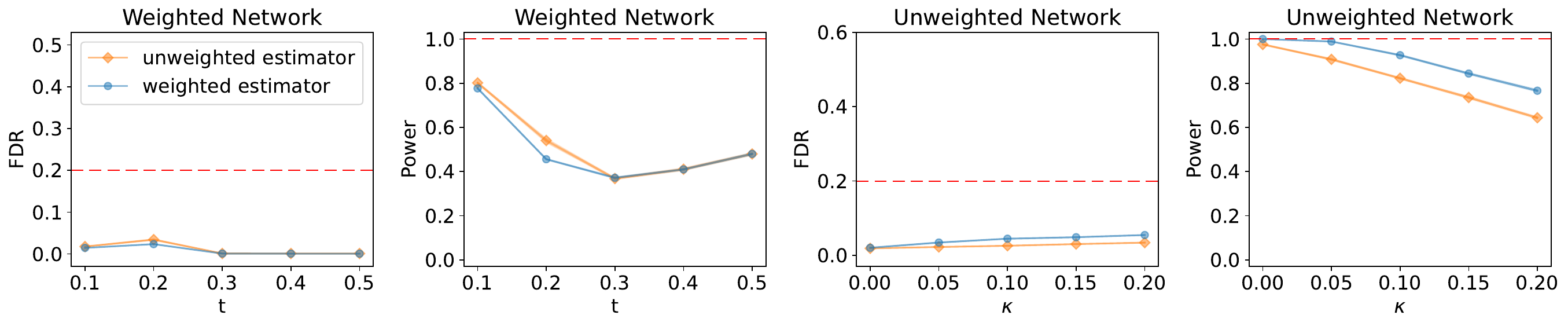}
    \caption{Comparisons of empirical FDRs and powers between weighted and unweighted estimators.}
    \label{fig::we-unwe-comparison}
\end{figure}

}

{\color{black}
\subsection{Sensitivity of {\tt clp} to the choice of $r_0$}

In our construction, the e-value for $j \in \mathcal{I}_{\text{test}}^{i_0}(k_0)$ contains the factor $|\mathcal{I}_{\text{test}}^{i_0}(k_0)|$ in the numerator, and the BH rejection set used in its denominator is also computed within this local test set. Therefore, the choice of $r_1$, which controls the upper bound on $|\mathcal{I}_{\text{test}}^{i_0}(k_0)|$, can affect the magnitude of the resulting e-values and hence the empirical performance of the final e-BH procedure. Specifically, a larger size of the local test set (larger $r_1$) may inflate e-values and potentially increase power, but it also reduces the size of each calibration subset $\mathcal{I}_{\text{calib}}^{i_0}(j_0)$, which may lead to more conservative conformal p-values in the BH procedure. 
Recall that in our {\tt clp} implementation, $r_1$ is determined by the calibration split size $|\Icalibi|$ and $r_0$ through $r_1 := \lfloor |\mathcal{I}_{\text{calib}}^{i_0}| / r_0 \rfloor$. Therefore, $r_1$ is not an entirely independent tuning parameter, but can be tuned indirectly through $r_0$, where a larger value of $r_0$ corresponds to a smaller $r_1$. 

To examine the sensitivity of {\tt clp} to the choice of $r_0$ (and hence $r_1$), we conducted additional experiments under Settings 1--3 from Section 3.1 of the main paper. We set $\kappa = 0.3$ and vary $r_0 \in \{10, 20, 30, 40\}$. The results are shown in Figure~\ref{fig::tune-r0}. 
\begin{figure}[h]
    \centering
    \includegraphics[width=0.9\textwidth]{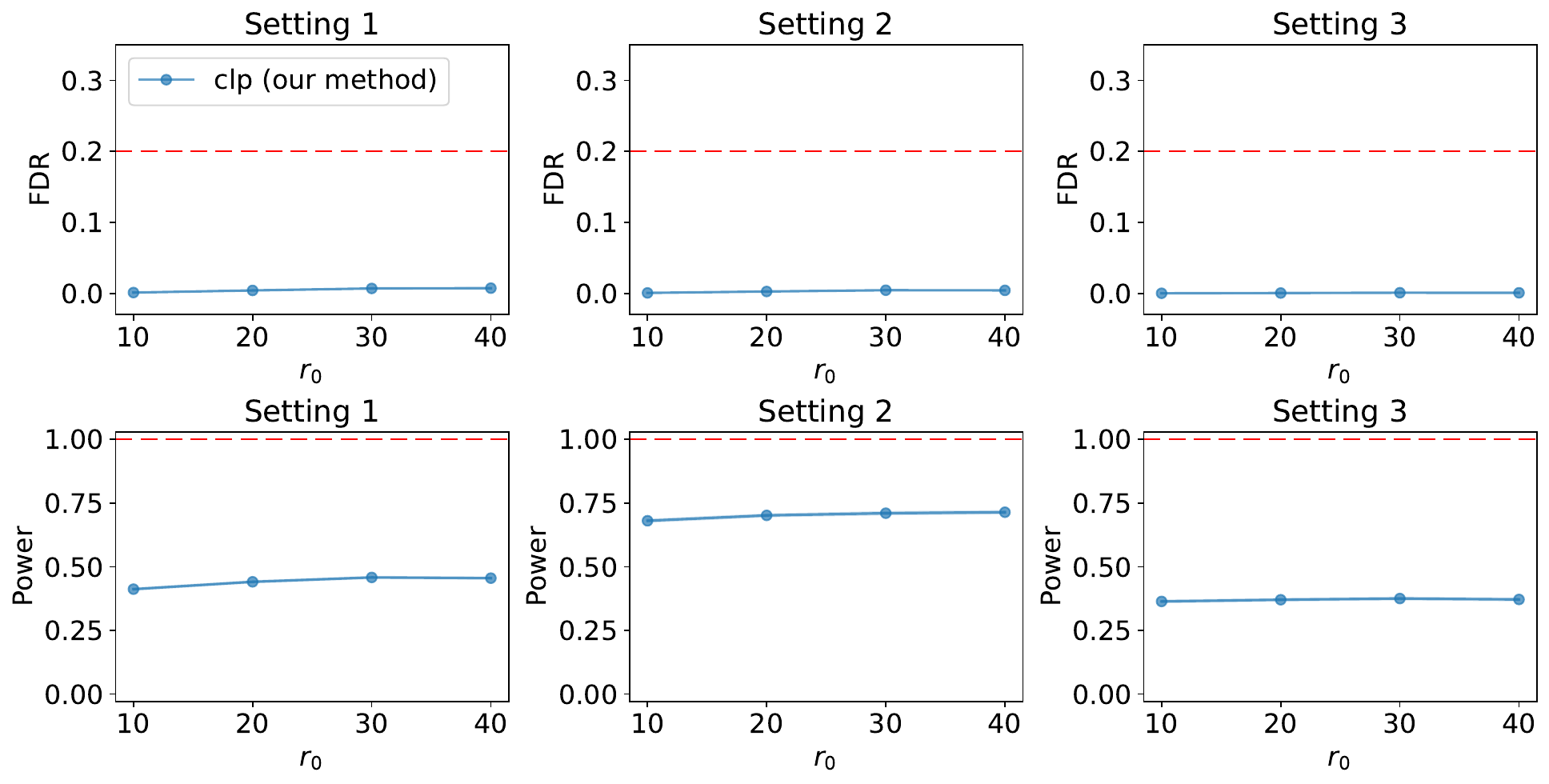}
    \caption{Sensitivity analysis of {\tt clp} with respect to $r_0$, which controls the local test set size $r_1 = \lfloor |\mathcal{I}_{\text{calib}}^{i_0}| / r_0 \rfloor$.}
    \label{fig::tune-r0}
\end{figure}
Across all three settings, the empirical FDR remains  below the nominal level for all values of $r_0$ considered, confirming that FDR control is robust to this choice. Additionally, the empirical power in Settings 1 and 2 increases slightly as $r_0$ increases, while in Setting 3 it remains nearly unchanged. Overall, these results suggest that the performance of our method is not sensitive to the local test set size over the considered range, and that $r_0 = 25$ as used in the main paper provides a reasonable default. 
}

\renewcommand\theequation{C.\arabic{equation}}
\setcounter{figure}{0}
\renewcommand\thefigure{C.\arabic{figure}}

\section{Additional real data analysis}
\label{supp::additional-real-data-analysis}

\begin{figure}[ht]
    \centering
    \includegraphics[width=0.95\textwidth]{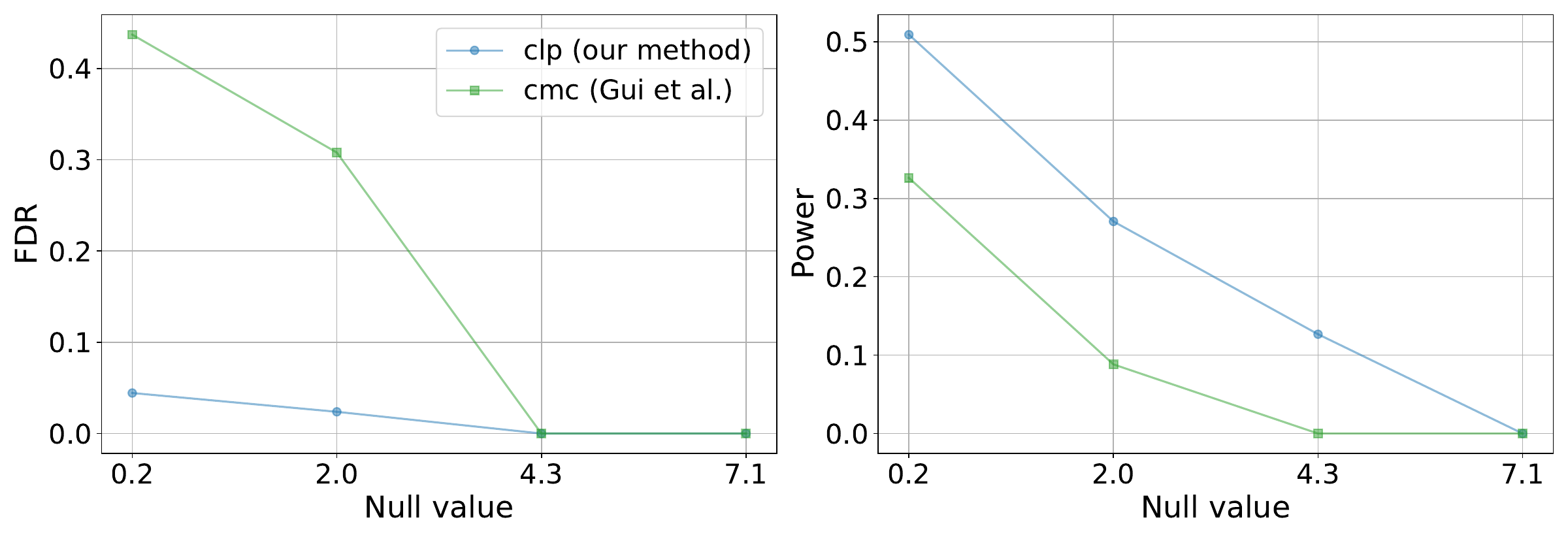}
    \vspace{-1em}
    \caption{FDR and power over $100$ independent runs for the international trade dataset with different null values.}
    \label{fig::trade-supp}
\end{figure}

In Section \ref{section::real-data}, we set $c_{i,j}=0.2$ and vary $\alpha \in \{0.1, 0.2, 0.3\}$.
In this section, we fix $\alpha=0.2$ and vary the value of $c_{i,j}$ to further demonstrate the performance of {\tt clp} and {\tt cmc}.
Specifically, we consider $c_{i,j} \in \{0.2, 2, 4.3, 7.1\}$, with the proportion of alternative hypotheses gradually decreasing to approximately 30\%.
The empirical FDR and power (averaging across $100$ independent runs) are summarised in Figure \ref{fig::trade-supp}.
As expected, {\tt clp} achieves valid FDR control across all configurations, while {\tt cmc} does not.
Moreover, as $c_{i,j}$ becomes larger, the power of both methods decreases.
As a result, {\tt clp} performs well in detecting whether trading activity exists between two countries but becomes less powerful when detecting higher volumes of bilateral trade.

\renewcommand\theequation{D.\arabic{equation}}
\renewcommand\thelemma{D.\arabic{lemma}}
\renewcommand\theproposition{D.\arabic{proposition}}
\setcounter{equation}{0}
\setcounter{lemma}{0}

\renewcommand\thetheorem{D.\arabic{theorem}}
\renewcommand\theequation{D.\arabic{equation}}
\setcounter{figure}{0}
\renewcommand\thefigure{D.\arabic{figure}}
\setcounter{algocf}{0} 
\renewcommand{\thealgocf}{D.\arabic{algocf}}

\section{FDR control for undirected and bipartite networks}
\label{supp-section::bipartitie-and-undirected}

In this section, we extend the proposed {\tt clp} framework to accommodate undirected and bipartite networks, elaborating on the modified version of Algorithm \ref{algorithm::sample-local-test}. Notice that once Algorithm \ref{algorithm::sample-local-test} and its properties are still valid to control the FDR for the local test  
as defined in Theorem \ref{thm:local-FDR-split}, we can directly incorporate it into Algorithm \ref{algorithm::BH-EBH-framework}  
and establish the FDR control across all missing links by applying the proof of Theorem \ref{thm:emp-fdr-global}. 

We start with the undirected networks.
Due to symmetry, we consider that information on the network and missing patterns is incorporated in the upper triangular part, \ie, $\{A_{i,j}\}$, $\{M_{i,j}\}$, for { $i<j$}. 
To facilitate the analysis, we write the extended version of $\bA$, $\bM$ as $\bA'$, $\bM'$, where
\begin{equation}
\label{eq:A-M-extend}
    A_{i,j}'=\left\{\begin{array}{cc}
       A_{i,j}  & \text{ if } { i<j} \\
        A_{j,i}  & \text{ if }  { i>j}
    \end{array}\right. ~~; \quad  M_{i,j}'=\left\{\begin{array}{cc}
       M_{i,j}  & \text{ if } { i<j} \\
        M_{j,i}  & \text{ if }  { i>j}
    \end{array}\right. .
\end{equation}
We now describe the FDR control for the local test 
in undirected networks in Alogrihtm \ref{algorithm::sample-local-test-und}.
\begin{algorithm}
\caption{Local test with FDR control for undirected network 
}
\label{algorithm::sample-local-test-und}
   
    \KwIn{Targeting row index $i_0$, set $\Itesti(k_0)$ where $j_0>i_0$ for each $j_0\in \Itesti(k_0)$.}
        Extend the $\bA$ and $\bM$ to $\bA'$ and $\bM'$ according to \eqref{eq:A-M-extend}; 

        Split the observed entries in row $i_0$ of $\bA'$ as $\Itraini$ ,$\Icalibi$ by missingness $\bM'$;

        Run Algorithm \ref{algorithm::sample-local-test} on $\bA'$, $\bM'$ for row $i_0$, set $\Itesti(k_0)$, given $\Itraini$ and $\Icalibi$;

    \KwOut{Rejection set $\cR_{\rm BH}^{i_0,k_0}(\alphaBH)$.} 
\end{algorithm}

\begin{proposition}[Conditional independence in undirected networks]\label{prop:emp-cond-ind-undirected} Suppose we run Algorithm \ref{algorithm::sample-local-test} on $\bA'$ and $\bM'$, with the test set $\Dtest$ in the upper triangular matrix.
Then, conditional on 
$\{\xi_j:j\in\Itraini\}$, $\{A_{i,j}:i,j\in\Itraini, { i<j}\}$ and the missing pattern $\bM$, the p-values in \eqref{eqn-def::emp-p-value} from  $\bA'$ and $\bM'$ are valid and independent within $\Itesti(k_0)$. 
\end{proposition}


\begin{figure}[H]
    \centering
    \includegraphics[width=\textwidth]{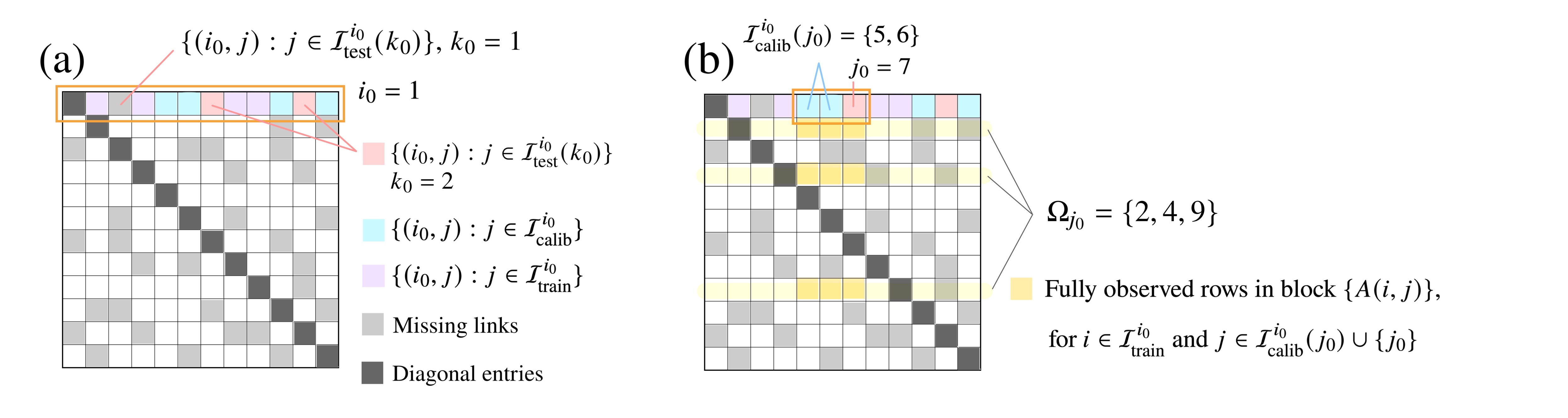}
    \vspace{-1.8em}
    \caption{
    Illustration of the splitting procedure in {\tt clp} under undirected network. 
    (a) Construction of $\Itraini$ and $\Icalibi$ for each $\Itesti(k_0)$. 
    (b) Construction of $\Omega_{j_0}$. 
    }
    \label{fig::clp-undirected-illus}
\end{figure}

Figure \ref{fig::clp-undirected-illus} provides an illustrative example of the splitting procedure in {\tt clp} for an undirected network. 
It is seen that the construction follows the same procedure as described in Section \ref{section::2-2} for directed networks.
Based on Proposition \ref{prop:emp-cond-ind-undirected}, the FDR control results below follows directly by the proof of Theorems \ref{thm:local-FDR-split}--\ref{thm:emp-fdr-global}.

\begin{theorem}[FDR control results for undirected networks] \label{thm:FDR-undirected}
Under the conditions of Proposition \ref{prop:emp-cond-ind-undirected}, in each local test set within row $i_0$: $\{H_{i_0,j_0}: j_0 \in \Itesti(k_0)\}$, the BH procedure using  $\wh{\mathfrak p}_{i_0,j_0},j_0 \in \Itesti(k_0)$ will lead to a valid FDR control: $\text{FDR}\le \alphaBH$ on this local test set. 
Moreover, the rejection set of {\tt clp} as the output of Algorithm \ref{algorithm::BH-EBH-framework} satisfies $\text{FDR} \leq \alphaeBH$ on the whole test set $\{H_{i,j}: i < j\in [n], M_{i,j}=1\}$, for any $\alphaBH, \alphaeBH \in (0,1)$.
\end{theorem}

We then turn to the bipartite networks, where $\bA, \bM\in \mathbb{R}^{n \times m}$ and the bipartite network $\bA$ follows the decomposition:
\begin{align}
    A_{i,j}\sim f(\xi_i,\zeta_j), \text{ ~for~ } i \in [n], \ j\in[m],
    \label{def::exch-model-bipartite}
\end{align}
where $\{\xi_i\}_{1\leq i\leq n}, \{\zeta_j\}_{1\leq j\leq m}\stackrel{\rm i.i.d.} \sim \mathrm{Uniform}[0,1]$. Notice that, the bipartite networks can be viewed as a $n\times m$ sub-network of the directed network \eqref{def::exch-model} in $\mathbb{R}^{ (n+m) \times (n+m)}$. Suppose that $\bA^{\text{full}}\in\mathbb{R}^{ (n+m) \times (n+m)}$ is a directed network generated by
$\big\{\{\xi_i\}_{1\leq i\leq n}, \{\zeta_j\}_{1\leq j\leq m}\big\}\stackrel{\rm i.i.d.} \sim \mathrm{Uniform}[0,1]$ following \eqref{def::exch-model}. Then, it is seen that $\bA=\bA^{\text{full}}[1:n,n+1:n+m]$ is a submatrix of the exchangeable $\bA^{\text{full}}$. Following this observation, we can show the exchangeablity of bipartite networks below.

\begin{lemma}[Exchangeablity in bipartite network]\label{lemma:exch-bipartite}
   Suppose $\bA\in \mathbb{R}^{n \times m}$ follows \eqref{def::exch-model-bipartite}, then for any permutation $\pi_1$ on $[n]$ and $\pi_2$ on $[m]$, we have 
   \begin{equation*}
       \{A_{i,j}\}_{i\in[n],j\in[m]}  \overset{d}{=}
    \{A_{\pi_1(i),\pi_2(j)}\}_{i\in[n],j\in[m]}.
   \end{equation*}
\end{lemma}
The poof of Lemma \ref{lemma:exch-bipartite} follows the property \eqref{def::exch-property} by noticing that any $\pi_1$ on $[n]$ and $\pi_2$ on $[m]$ can be combined as a unique $ \pi_{\text{full}}$ acting on $[n+m]$ with 
        \begin{equation*}
           \pi_{\text{full}}(i)=  \left\{ \begin{array}{cl}
             \pi_1(i) & \text{when } i\in [n], \\
            \pi_2(i-n)+n & \text{when } i \in[n+m] \backslash [n]. \\
                    \end{array}\right. 
        \end{equation*}
We now modify our Algorithm \ref{algorithm::sample-local-test} to the following version by substituting $\Itraini$ in step 3 by $[n]{\backslash \{i_0\}}$:

\begin{algorithm}[h]
\caption{Local test with FDR control for bipartite network 
}
\label{algorithm::sample-local-test-bipartite}
   
    \KwIn{Targeting row index $i_0$, set $\Itesti(k_0)$, $\Icalibi$, $\Itesti$ and
    calibration size $r_0 \ge C/\alphaBH $.}

        Split $\Icalibi$ into $|\Itesti(k_0)|$ parts evenly, \textit{i.e.}, $\{\Icalibi(j_0)\}_{j_0}$, for each $j_0\in \Itesti(k_0)$\;

       \For{$j_0\in \Itesti(k_0)$ }{
        
        Find the fully observed rows in the block $\{A(i,j)\}$, for $i\in [n]{\backslash \{i_0\}}$ and $j\in \Icalibi(j_0)\cup\{j_0\}$, which we denote by $\Omega_{j_0}$\;

        Calculate $\wh A_{i_0,j}$ for $j \in\Icalibi(j_0)\cup\{j_0\}$ using \eqref{def::d_{j_1,j_2,j}}--\eqref{method::A-estimate}\;
        Compute p-value for $j_0$ using \eqref{eqn-def::emp-p-value}\;
       }
       Run BH procedure for p-values in $\Itesti(k_0)$ and record the rejection results $\cR_{\rm BH}^{i_0,k_0}(\alphaBH)$\; 

    \KwOut{Rejection set $\cR_{\rm BH}^{i_0,k_0}(\alphaBH)$.} 
\end{algorithm}
By the construction of $\Omega_{j_0}$, the submatrix $[\bA\circ \bM^c]_{\Omega_{j_0},\Icalibi(j_0)\cup\{j_0\} }$ is always fully observed. 
According to Lemma \ref{lemma:exch-bipartite}, we can show the following exchangeability property within the submatrix.
\begin{proposition} 
\label{prop::col-index-exch-v2}
    Conditional on the missing pattern $\bM$ and $\{\xi_i\}_{i\in\Omega_{j_0}}$, $\{\zeta_j\}_{j\in \Itraini}$, the submatrix $ [\bA\circ \bM^c]_{\Omega_{j_0},\Icalibi(j_0)\cup\{j_0\} }$ is exchangeable with respect to the columns indices, \ie,
        \begin{equation*}
        \{[\bA\circ \bM^c]_{i,j}\}_{i\in \Omega_{j_0}, j\in \Icalibi(j_0)\cup\{j_0\}} \overset{d}{=} \{[\bA\circ \bM^c]_{i,\pi(j)} \}_{i\in \Omega_{j_0}, j\in \Icalibi(j_0)\cup\{j_0\}} 
    \end{equation*}
    where  $\pi(\cdot)$ is any permutation of the index set $\Icalibi(j_0)\cup\{j_0\}$.
\end{proposition}


\begin{figure}[H]
    \centering
    \includegraphics[width=\textwidth]{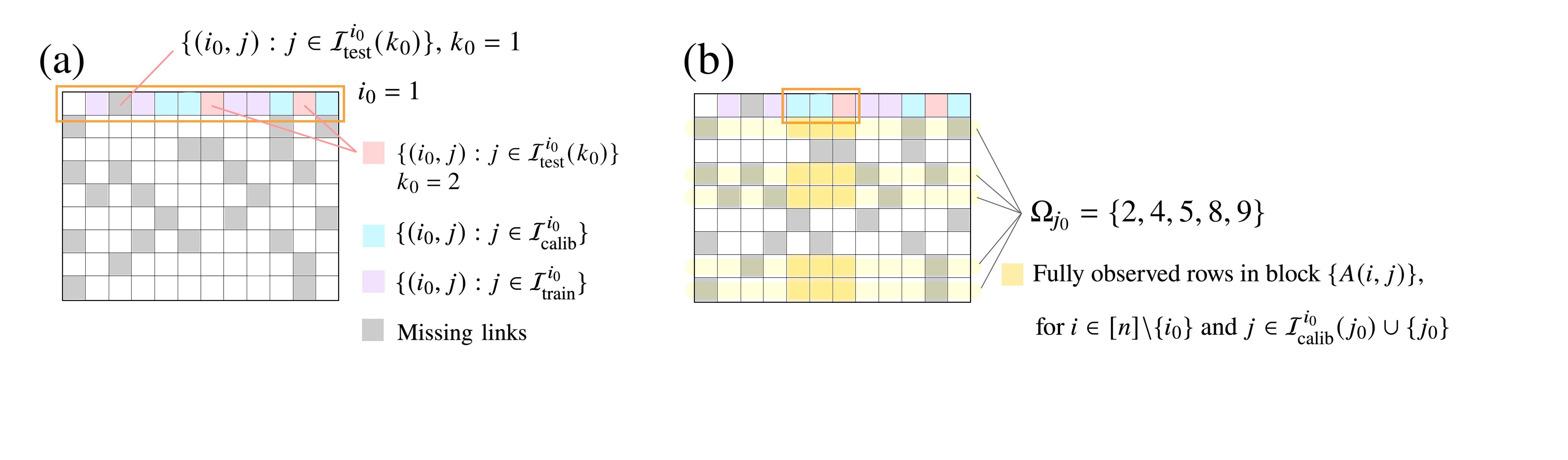}
    \vspace{-2.5em}
    \caption{
    Illustration of the splitting procedure in {\tt clp} under bipartite network. 
    (a) Construction of $\Itraini$ and $\Icalibi$ for each $\Itesti(k_0)$. 
    (b) Construction of $\Omega_{j_0}$. 
    }
    \label{fig::clp-bipartite-illus}
\end{figure}

An example illustrating the splitting procedure in {\tt clp} for bipartite networks is shown in Figure \ref{fig::clp-bipartite-illus}.
Now we are ready to construct the conformal p-values.
We can analogously define $\Omega_{j_1,j_2,j}$ as in the main paper.
By construction, for every given $(j_2, j)$,  where $j_2 \in \Itraini$, all $j_1\in \Icalibi(j_0) \cup \{j_0\}$ share a common $\Omega_{j_1,j_2,j}$.
By Proposition \ref{prop::col-index-exch-v2}, conditional on the training data $\{\xi_i\}_{i\in\Omega_{j_0}}$, $\{\zeta_j\}_{j\in\Itraini}$, training block $\{A_{i,j}\}_{i\in\Omega_{j_0},j\in\Itraini}$ and missing pattern $\bM$, it is seen that for each $j_2\in \Itraini$, we have 
\allowdisplaybreaks
\begin{equation*}
    \begin{aligned}
           \wh{d}(\pi(j_1),j_2)
    =&~
    \sum_{j \in \Itraini \backslash \{j_2\}} 
        \frac{
            |\langle A_{\Omega_{\pi(j_1),j_2,j},\pi(j_1)} - A_{\Omega_{\pi(j_1),j_2,j},j_2}, A_{\Omega_{\pi(j_1),j_2,j},j} \rangle|
        }
        {
            \big( |\Itraini| - 1 \big) |\Omega_{\pi(j_1),j_2,j}|
        } \\
         \overset{d}{=} & \sum_{j \in \Itraini \backslash \{j_2\}} 
        \frac{
            |\langle A_{\Omega_{j_1,j_2,j},j_1} - A_{\Omega_{j_1,j_2,j},j_2}, A_{\Omega_{j_1,j_2,j},j} \rangle|
        }
        {
            \big( |\Itraini| - 1 \big) |\Omega_{j_1,j_2,j}|
        }\\
        =&~~ \wh{d}(j_1,j_2).
    \end{aligned}
\end{equation*}
The second equality above is due to the the definition of inner product and Proposition \ref{prop::col-index-exch-v2}. This directly gives the following proposition, which can be justified by applying the proof of Proposition \ref{prop:emp-cond-ind}. 

\begin{proposition}[Conditional independence between empirical p-values]\label{prop:emp-cond-ind-bipartite}
Conditional on the training data $\{\xi_i\}_{i\in\Omega_{j_0}}$, $\{\zeta_j\}_{j\in\Itraini}$, training block $\{A_{i,j}\}_{i\in\Omega_{j_0},j\in\Itraini}$ and missing pattern $\bM$, the p-values in \eqref{eqn-def::emp-p-value} are valid and independent within $\Itesti(k_0)$. 
\end{proposition}

Analogous to Theorem \ref{thm:FDR-undirected}, according to Proposition \ref{prop:emp-cond-ind-bipartite}, we obtain the following FDR control results.

\begin{theorem}[FDR control results for bipartite networks]
\label{thm:FDR-bipartite}
Under the conditions of Proposition \ref{prop:emp-cond-ind-bipartite}, in each local test set within row $i_0$: $\{H_{i_0,j_0}: j_0 \in \Itesti(k_0)\}$, the BH procedure using $\wh{\mathfrak p}_{i_0,j_0},j_0 \in \Itesti(k_0)$ leads to a valid FDR control: $\text{FDR}\le \alphaBH$ on this local test set. 
Moreover, the rejection set of {\tt clp} by inserting Algorithm \ref{algorithm::sample-local-test-bipartite} to Algorithm \ref{algorithm::BH-EBH-framework} satisfies $\text{FDR} \leq \alphaeBH$ on the test set $\{H_{i,j}: i \in [n], j\in [m], M_{i,j}=1\}$, for any $\alphaBH, \alphaeBH \in (0,1)$.
\end{theorem}

\renewcommand\thetheorem{E.\arabic{theorem}}
\renewcommand\theequation{E.\arabic{equation}}
\setcounter{figure}{0}
\renewcommand\thefigure{E.\arabic{figure}}
\setcounter{algocf}{0} 
\renewcommand{\thealgocf}{E.\arabic{algocf}}

{
\section{Additional theoretical results}
\label{supp::Bern-M}
We now show the existences of $\Omega_{j_0}$ for matching that leads to non-trivial rejections of our algorithm.
The proposition below guarantees that, when $M_{i,j}$ are independent Bernoulli$(q_{i,j})$ random variables with unknown and potentially heterogeneous missing probabilities $q_{i,j}$, there exists a sufficiently large number of fully observed rows to construct $\bA_{\Omega_{j_0}\times (\Icalibi(j_0) \cup \{j_0\})}$, provided the missing rates satisfy the mild condition $q_{i,j} \le c_0 < 1$.

\begin{proposition}[Number of fully observed rows in columns $\Icalibi(j_0) \cup \{j_0\}$]\label{prop:num-match-rows}
  Suppose the missing probabilities $q_{i,j}\le c_0<1$ for $i,j \in [n]$, and there exists a constant $C_1>0$ such that $|\Itraini| = C_1 n$, and $|\Icalibi(j_0)|= r_0$. Then we have $|\Omega_{j_0}|\ge {C_1}(1-c_0)^{r_0+1} n/2$ with probability at least $1-\exp\{-n C_1 (1-c_0)^{2r_0+2}/16\}$.
\end{proposition}


\begin{remark}[Splitting procedure]
    Note that when the missing rate $q_{i,j}\in [c,c']$ for some $c,c' \in [0,1)$, we typically have $|\Itesti|\asymp n $. 
    If we split the train and calibration sets equally, we have $
    |\Itraini|\asymp |\Icalibi|\asymp |\Itesti|\asymp n$. This implies that in multiple splitting, we can only demand the size of each split  $|\Icalibi(j_0)|\approx |\Icalibi|/|\Itesti| \asymp 1$, \ie, $|\Icalibi(j_0)|$ is at a constant level. Such a constant size will lead to high randomness of multiple testing, but on the other hand, it can guarantee sufficient large number of fully observed rows $\Omega_{j_0}$ and thus maintain the estimation error as well as the power of selection. Moreover, when the missing values are scarce, we have small $|\Itesti|$. In this case, the splitting on $\Itesti$ is not required, \ie, $\Itesti(k_0)=\Itesti$. Instead, we can split $\Icalibi$ into more than $|\Itesti|$ sets such that the size of each subset is at a moderate level (to ensure enough fully observed rows in $\Omega_{j_0}$), and apply e-BH using part of the subsets.
    
    
\end{remark}

Moreover, we show that under structured missingness patterns, such as block or staggered missingness \citep{athey2021matrix,xiong2023large}, there remain sufficiently many fully observed rows for each $(i_0,j_0)$ under mild regularity conditions. We take block missingness as an illustrative example; the argument for staggered missingness follows analogously.
\begin{proposition}[Fully observed rows under block missingness]\label{prop:num-match-rows-block}
Suppose the missingness pattern satisfies $q_{i,j}=1$ for $i \ge i_n$ and $j \ge j_n$, and $q_{i,j}=0$ otherwise, where $i_n/n \ge 1 - c_1$ and $j_n/n \ge 1 - c_1$ for some constant $c_1 > 0$. Then, if choosing $|\Itraini| = c_2 n$ for some constant $c_2 < c_1$, we have $|\Omega_{j_0}| = |\Itraini| = c_2 n$ for each $j_0$ and for any choice of $\Icalibi(j_0)$.
\end{proposition}
}

\end{document}